\documentclass[journal]{IEEEtran}

\IEEEoverridecommandlockouts
\makeatletter
\def\ps@headings{%
\def\@oddhead{\mbox{}\scriptsize\rightmark \hfil \thepage}%
\def\@evenhead{\scriptsize\thepage \hfil \leftmark\mbox{}}%
\def\@oddfoot{}%
\def\@evenfoot{}}
\makeatother
\usepackage{subfigure}
\usepackage{colortbl}
\usepackage{bm}

\usepackage{graphicx}  
\usepackage{url}       

\usepackage{amsmath}   
\usepackage{cite}

\usepackage{amsfonts,amssymb}
\usepackage{cite}
\usepackage{stfloats}

\usepackage{cases}
\usepackage{algorithm}
\usepackage{multirow}
\usepackage{algorithmic}
\usepackage{stfloats}
\usepackage{epstopdf}

\makeatletter

\newcommand{\Rmnum}[1]{\expandafter\@slowromancap\romannumeral #1@}
\makeatother

\UseRawInputEncoding

\newtheorem{theorem}{{Theorem}}

\newcommand{\ls}[1]
    {\dimen0=\fontdimen6\the\font
     \lineskip=#1\dimen0
     \advance\lineskip.5\fontdimen5\the\font
     \advance\lineskip-\dimen0
     \lineskiplimit=.9\lineskip
     \baselineskip=\lineskip
     \advance\baselineskip\dimen0
     \normallineskip\lineskip
     \normallineskiplimit\lineskiplimit
     \normalbaselineskip\baselineskip
     \ignorespaces
    }

\begin{document}

\title{Dynamic Energy-Saving Design for Double-Faced Active RIS Assisted Communications with Perfect/Imperfect CSI}
\vspace{10pt}

\author{\IEEEauthorblockN{Yang Cao, \emph{Graduate Student Member}, \emph{IEEE}, Wenchi Cheng, \emph{Senior Member}, \emph{IEEE}, Jingqing Wang, \emph{Member}, \emph{IEEE}, and Wei Zhang, \emph{Fellow}, \emph{IEEE}}\\[0.2cm]
	\vspace{-5pt}
	
	
	\vspace{-25pt}
	
	\thanks{
		
		Yang Cao, Wenchi Cheng and Jingqing Wang are with the State Key Laboratory of Integrated Services Networks, Xidian University, Xi'an,
		710071, China (e-mails: caoyang@stu.xidian.edu.cn, wccheng@xidian.edu.cn, wangjingqing00@gmail.com).
		
		Wei Zhang is with the School of Electrical Engineering and Telecommunications, University of New South Wales, Sydney,
		NSW 2052, Australia (e-mail: w.zhang@unsw.edu.au).}
}

\maketitle

\begin{abstract}
  Although the emerging reconfigurable intelligent surface (RIS) paves a new way for next-generation wireless communications, it suffers from inherent flaws, i.e., double-fading attenuation effects and half-space coverage limitations. The state-of-the-art double-face active (DFA)-RIS architecture is proposed for significantly amplifying and transmitting incident signals in full-space. Despite the efficacy of DFA-RIS in mitigating the aforementioned flaws, its potential drawback is that the complex active hardware also incurs intolerable energy consumption.
  To overcome this drawback, in this paper we propose a novel dynamic energy-saving design for the DFA-RIS, called the sub-array based DFA-RIS architecture. This architecture divides the DFA-RIS into multiple sub-arrays, where the signal amplification function in each sub-array can be activated/deactivated dynamically and flexibly. Utilizing the above architecture, we develop the joint optimization scheme based on transmit beamforming, DFA-RIS configuration, and reflection amplifier (RA) operating pattern to maximize the energy efficiency (EE) of the DFA-RIS assisted multiuser MISO system considering the perfect/imperfect channel state information (CSI) case. Then, the penalty dual decomposition (PDD) based alternating optimization (AO) algorithm and the constrained stochastic majorization-minimization (CSMM) based AO algorithm address non-convex problems in the perfect/imperfect CSI case, respectively. Simulation results verified that our proposed sub-array based DFA-RIS architecture can benefit the EE of the system more than other RIS architectures.

\end{abstract}



\section{Introduction}
 \IEEEPARstart{I}{n} recent years, worldwide efforts have been carried out to explore potential new technologies and applications for next-generation wireless communication systems \cite{zongshu1,OAM-NFC,FullDuplex}. The emerging reconfigurable intelligent surface (RIS) technology \cite{Self-Interference} has been favored by both academia and industry, and has been envisioned to pave a new way for the development of next-generation communication systems.
 As a physically structured surface, the RIS consists of numerous energy-efficient, cost-effective unit reflecting elements (REs) and a controller. The REs of RIS can be delicately adjusted to modify the phase shift and/or amplitude of the incident signal to purposefully reshape the beam of the signal. Based on this signal reshaping capability, the RIS can effectively create a tunable communication environment for improving communication performance metrics in terms of achievable rate, power consumption, secrecy rate, etc. The great potential of RIS has been profoundly investigated in recent studies \cite{Max_R,Joint-Active,MISO-SEC,Cao1}. For instance, the authors of \cite{Max_R} solved the problem of the achievable average sum-rate maximization in RIS-assisted multi-user systems, showing that RIS can contribute effectively to improving system performance. The authors of \cite{Joint-Active} investigated a single-cell RIS-assisted wireless system and addressed the problem of minimizing the total transmit power by jointly optimizing the transmit beamforming and the reflected beamforming at RIS.  The authors of \cite{MISO-SEC} verified that RIS can significantly improve the secrecy rate in multiple-input-single-output (MISO) systems.  The authors of \cite{Cao1} investigated a low-complexity multi-RIS assisted anti-jamming optimization strategy, revealing that RIS can effectively contain jamming.

 
 However, some of the inherent flaws of the traditional passive RIS have been gradually revealed in the ongoing research of RIS. First and foremost, due to hardware limitations, conventional RIS is limited to only reflective or transmissive transmission of incident signals, thus serving the user only in a half-space. In order to achieve full spatial coverage, a simultaneous transmitting and
 reflecting RIS (STAR-RIS) structure has been proposed \cite{STAR_RIS1}. In contrast to the conventional RIS, this novel structure reflects and transmits the incident signal simultaneously at the RIS, with the transmitted portion of the signal penetrating the surface and propagating into the back half of space. The authors of \cite{STAR_RIS1} introduced the basic signaling model of STAR-RIS and proposed three operating mode protocols for STAR-RIS. The authors of \cite{STAR_RIS2} investigated the channel model of STAR-RIS in the near-field and far-field cases and further analyzed the corresponding diversity gains. 
 

 Unfortunately, the other inherent flaw, double-fading attenuation, limits the achievable performance of STAR-RIS and conventional passive RIS. Double fading means that a cascaded channel constructed by the RIS subjects the passing signal to two large-scale fadings, resulting in a smaller performance gain compared to fading without the RIS.
 In order to compensate for the double-fading attenuation, the active RIS architecture has been proposed \cite{Active_RIS1}. The unique structure of the active RIS features an additional amplifier for each RE, which makes it capable of simultaneously adjusting the phase and amplitude of weak incident signals. The advantage of active RIS over relays or repeaters is that sufficient signal amplification can be obtained without the benefit of a high consumption radio frequency (RF) link. Therefore, active RIS can compensate for double-fading attenuation with low hardware costs. The authors of \cite{Active_RIS2} demonstrated that active RIS can achieve rate gains tens of times higher than passive RIS in the system. However, the architecture of the active RIS also imposes a limitation, confining its serviceability to users within half-space, resulting in the inability to perform full-space communication.
 
 Both architectures, STAR-RIS, and active RIS, have their own limitations, i.e., they either suffer from the double-fading attenuation or are constrained to achieving half-space coverage. It is difficult for them to address both these challenges simultaneously. In order to fully unleash the potential and advantages of STAR-RIS and active RIS, the authors of \cite{DFA-RIS1} proposed novel double-faced active (DFA)-RIS architectures, which can amplify the incident signal while allowing the signal to penetrate the surface to reach the backside. In addition, DFA-RIS is equipped with tunable power division units (PDUs) to allow fine power distribution between reflection and transmission. The authors of \cite{DFA-RIS2} and \cite{DFA-RIS3} investigated the benefits of DFA-RIS-assisted multi-user communication systems with respect to the improved spectral efficiency and enhanced security, respectively. Although the DFA-RIS architecture has been demonstrated to offer superior performance gains compared to the other RIS architectures mentioned above, simultaneously incorporating reflection
 amplifiers (RAs) and PDUs into the system leads to an increase in both the energy consumption and hardware overhead of the DFA-RIS. 
 
 In order to reduce the energy consumption of DFA-RIS operation, a possible solution is to finely manipulate the process of DFA-RIS to modulate the incident signal. Motivated by the aforementioned considerations, in this paper we propose a flexible dynamic energy-saving design for DFA-RIS to reduce the power consumption and improve the energy efficiency (EE) of the DFA-RIS based system. This design is a sub-array based DFA-RIS architecture that divides RE pairs with their corresponding connected RAs and PDUs into multiple sub-arrays, where the power supply of the RAs in each sub-array is controlled by a switch that can be switched on/off on demand. In order to validate the performance of the above architecture, we formulate optimization problems to maximize the EE of a sub-array based DFA-RIS assisted multiuser MISO system for both perfect and imperfect channel state information (CSI) cases by jointly optimizing the transmit beamforming,  DFA-RIS configuration, and RA operating pattern. For perfect CSI setup, we adopt the penalty dual decomposition (PDD) based alternating optimization (AO) algorithm to address the non-convex problem to maximize the EE of system. In conjunction with the algorithm for the perfect CSI case, we develop the constrained stochastic majorization-minimization (CSMM) based AO algorithm to efficiently find a suboptimal solution for the imperfect CSI case. Finally, numerical results indicate that the EE of the system can benefit more from our proposed sub-array based DFA-RIS scheme compared to other schemes.

 The rest of this paper is organized as follows. Section~\ref{sec:System_model} introduces the dynamic energy-saving design for DFA-RIS and system model. Section~\ref{sec:Power_Consumption_Problem_formulation} investigates the power consumption model of the DFA-RIS, and formulates the non-convex  problem to maximize the EE in the perfect/imperfect CSI cases. In Section~\ref{sec:Alternating_optimization_perfect_CSI}, the PDD-Based AO algorithm is proposed to solve the non-convex problem for the perfect CSI case. Section~\ref{sec:Alternating_optimization_imperfect_CSI} investigates the CSMM-Based AO algorithm to solve the non-convex problem for the imperfect CSI case.
 In Section~\ref{sec:simulation}, the numerical results are presented to evaluate the system performance. This paper concludes with Section~\ref{sec:Conclusion}.
 
 \textit{Notations}: Throughout this paper, boldface straight lowercase and uppercase letters denote vectors and matrices, respectively. The conjugate transpose and transpose are denoted by $(\cdot)^H$, and $(\cdot)^T$, respectively. The notations   $\lVert\cdot\rVert$ and $\lvert\cdot\rvert$ denote the norm of a matrix and the absolute value of a scalar. $\boldsymbol{\rm I}$ is an identity matrix with proper dimension. The distribution of complex Gaussian random vector with mean 0 and variance $\sigma^2$ is represented by $\mathcal{CN}(0,\sigma^2)$. $\mbox{Re}\left\{\cdot\right\}$, $\mbox{Diag}(\cdot)$, $\mbox{blkDiag}(\cdot)$, and $\mbox{arg}\left(\cdot\right)$ represent the real part of a complex variable, a diagonal matrix, chunked diagonal matrix, and the extraction of phase information, respectively.

\begin{figure*}[t]
	\vspace{-18pt}
	\centering
	\includegraphics[scale=0.080]{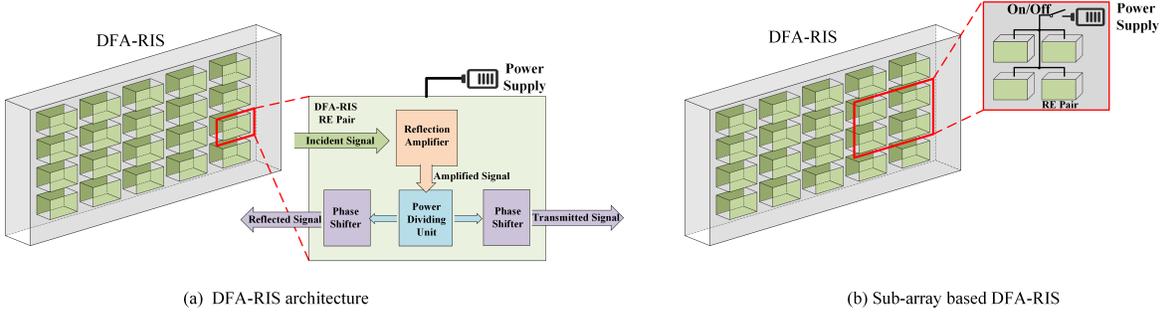}
	\vspace{-10pt}
	\caption{Illustration of the DFA-RIS  architecture and the dynamic energy-saving design for DFA-RIS: A sub-array based DFA-RIS architecture.}
	\vspace{-13pt}
	\label{fig:shiyitu}
\end{figure*}

 \section{The Dynamic Energy-saving Design for DFA-RIS And System Model}
 \label{sec:System_model}

 \subsection{The Dynamic Energy-Saving Design for DFA-RIS}
 \label{sec:Dynamic_Energy-saving}
 DFA-RIS has been recently proposed to overcome the double-fading attenuation and to fully cover the whole space. The architecture of DFA-RIS is illustrated in Fig.~\ref{fig:shiyitu}{(a)}, where reflecting element (RE) arrays are deployed on both sides of the plate with their REs being aligned. Each pair of aligned REs located on opposite faces are connected to a tunable PDU and a reflection amplifier (RA). The incident signal amplified by the RA is split into two portions in the PDU and then fed into the opposite phase shifter for phase adjustment and output. Thus, DFA-RIS can achieve amplification, phase modulation, and simultaneous bi-directional transmission (reflection and transmission) of the incident signal.

 However, to realize the comprehensive functions mentioned above, DFA-RIS requires a plenty of RAs to operate, resulting in enormous power consumption. This shortcoming motivates us to propose the dynamic energy-saving design for DFA-RIS. Specifically, we consider appropriately turning off a portion of the RAs to reduce energy consumption. In this regard, we propose a sub-array based DFA-RIS architecture, where the power supply of RAs in each sub-array is controlled by a switch, which can be switched on/off on demand, as shown in Fig.~\ref{fig:shiyitu}{(b)}. The introduction of the architecture of DFA-RIS based on sub-array leads to a significant reduction in the number of operational RAs, which can enhance the EE of the system. But at the same time, the reduction of the degree of freedom (DoF) of DFA-RIS entails certain performance losses.
 
 \subsection{Signal Model of DFA-RIS Based on Sub-Array}
 \label{sec:Signal_Model}
 
 Utilizing the DFA-RIS architecture based on sub-array, we develop the signal model of DFA-RIS. Specifically, with $M$ REs deployed on each side of the DFA-RIS, there are a total of $M$ pairs of aligned REs. Let the set of pairs of REs be defined by $\mathcal{M}=\{1,2,\cdots,M\}$. We divide $M$ pairs of REs into $L$ sub-arrays, where the operating state of the RAs in each sub-array is uniformly controlled by a switch. Each sub-array contains $M_{sub}=M/L$ pairs of REs. Let the set of sub-arrays be defined by $\mathcal{L}=\{1,2,\cdots,L\}$. Let $\boldsymbol {\rm x}_{in}$ be denoted as the incident signal in the DFA-RIS, hence the reflected and transmitted signals (denoted by $\boldsymbol{\rm y}_{R}$ and $\boldsymbol{\rm y}_{T}$) at the output of the DFA-RIS are further modeled as respectively
 \begin{small}
 	\begin{align}
 		\boldsymbol{\rm y}_{R} = \boldsymbol{\rm \Theta}_{R}\boldsymbol{\rm D}_{R}\boldsymbol{\rm A}(\boldsymbol {\rm x}_{in}+\boldsymbol {\rm n}_{\scriptscriptstyle R\hspace{-0.2mm}I\hspace{-0.2mm}S}),\\
 		\boldsymbol{\rm y}_{T} = \boldsymbol{\rm \Theta}_{T}\boldsymbol{\rm D}_{T}\boldsymbol{\rm A}(\boldsymbol {\rm x}_{in}+\boldsymbol{\rm n}_{\scriptscriptstyle R\hspace{-0.2mm}I\hspace{-0.2mm}S}),
 	\end{align} 
 \end{small}
 where $\boldsymbol {\rm \Theta}_R= {\rm Diag}(\boldsymbol{\rm {\theta}}_R)$, $\boldsymbol {\rm \Theta}_T= {\rm Diag}(\boldsymbol{\rm {\theta}}_T)$, and $\boldsymbol{n}_{\scriptscriptstyle R\hspace{-0.2mm}I\hspace{-0.2mm}S}\sim\mathcal{CN}(0,\sigma^2_{\scriptscriptstyle R\hspace{-0.2mm}I\hspace{-0.2mm}S}\boldsymbol{\rm I}_M)$ denote the phase shift matrices of the reflected and transmitted signals, and the additive complex Gaussian noise at the DFA-RIS, respectively, where $\boldsymbol{\rm {\theta}}_R = [\theta_{R,1},\theta_{R,2},\cdots,\theta_{R,M}]^T$ and $\boldsymbol{\rm {\theta}}_T = [\theta_{T,1},\theta_{T,2},\cdots,\theta_{T,M}]^T$. The notations $\theta_{R,m}=e^{j\phi_{R,m}}$ and $\theta_{T,m}=e^{j\phi_{T,m}}$ represent phase-shifts of the reflected and transmitted signals on the $m$th pair of REs. Besides, $\boldsymbol{\rm D}_{R}={\rm Diag}(\boldsymbol{\rm {\beta}})$ and $\boldsymbol{\rm D}_{R}={\rm Diag}(\sqrt{1-\boldsymbol{\rm {\beta}}^2})$ indicate the corresponding power splitting parameter matrices, where $\boldsymbol{\rm{\beta}}=[\beta_{1},\beta_{2},\cdots,\beta_{M}]^T$. The notations $\boldsymbol{\rm A}={\rm blkDiag}([\boldsymbol{\rm A}_{1},\boldsymbol{\rm A}_{2},\cdots,\boldsymbol{\rm A}_{L}])$ represents amplifying coefficient matrix, where $\boldsymbol{\rm A}_{l}={\rm Diag}(\boldsymbol{\rm \alpha}_{l}), \forall l\in \mathcal{L}$ with $\boldsymbol{\rm \alpha}_{l}=[\alpha_{(l-1)M_{sub}+1},\cdots,\alpha_{lM_{sub}}]^T$ being the amplifying coefficient  of the $l$th sub-array. Besides, $\alpha_{m}\in [0,\alpha_{\max}], (\alpha_{\max}\ge 1)$ indicates the amplifying coefficient of the $m$th pair of REs with $\alpha_{\max}$ being the maximum amplified coefficient. In addition, based on the sub-array based architecture of DFA-RIS, we define the set of sub-arrays with activated RAs as $\mathcal{L}_s \subseteq \mathcal{L}$, such that $\boldsymbol{\rm \alpha}_{l}\neq\boldsymbol{\rm 1}_{M_{sub}}$ if $l\in\mathcal{L}_s$, and $\boldsymbol{\rm \alpha}_{l}=\boldsymbol{\rm 1}_{M_{sub}}$ if $l\notin\mathcal{L}_s$, where $\boldsymbol{\rm 1}_{M_{sub}}$ represents the all-one $M_{sub}\times 1$ column vector.

 \subsection{System Model}
 \label{sec:System_Model}
 
 In this paper, we consider a DFA-RIS assisted multiuser MISO system, where a BS equipped with $N$ antennas provides communication services to $K$ single-antenna users with the assistance of DFA-RIS. Let the set of users be denoted by $\mathcal{K}=\{1,2,\cdots,K\}$. Further, the sets of users that fall in the reflective and transmissive areas of the DFA-RIS are set to be $\mathcal{K_R}=\{1,\cdots,K_R\}$ and $\mathcal{K_T}=\{K_R+1,\cdots,K_R+K_T\}$, respectively, where $K = K_R + K_T$. Let  $\boldsymbol{\rm G} \in \mathbb{C}^{M\times N}$, $\boldsymbol{\rm h}_{d,k}^H \in \mathbb{C}^{1\times N}$, and $\boldsymbol {\rm h}_{r,k}^H \in \mathbb{C}^{1\times M}$ represent the channel coefficients between the BS and the DFA-RIS, between the BS and the $k$th user, and between the DFA-RIS and the $k$th user, respectively. The quasi-static flat-fading model is assumed for all the aforementioned channels. 
 
 The transmitted signal vector, denoted by $\boldsymbol{\rm {x}}$, at the BS is given by
 \begin{small}
 	\begin{align}
 		\boldsymbol{\rm {x}} = \sum_{k=1}^{K}\boldsymbol{\rm {w}}_ks_k,
 	\end{align}
 \end{small}
 where $\boldsymbol{\rm {w}}_k\in \mathbb{C}^{N\times 1}$ and $s_k$ are the transmit beamforming vector and the transmit symbol for the $k$th user, respectively, where $s_k$ is modeled as a zero mean and unit variance random variable with independent and identically distributed (i.i.d.), and $s_k\sim\mathcal{CN}(0,1)$. To simplify notations, we introduce the notation $\varrho(k)$ to denote the reflective $(\varrho(k) = R)$ and transmissive $(\varrho(k) = T)$ regions. $\varrho(k)\in \mathcal{U}\triangleq\{R,T\}$, and $\bar{\varrho}(k)=\mathcal{U}/\varrho(k)$, such that if $\varrho(k) = R$, $\bar{\varrho}(k)=T$, and vice versa. Thus, we have $\varrho(k)=R, \forall k\in \mathcal{K_R}$, and $\varrho(k)=T, \forall k\in \mathcal{K_T}$. Then, based on the signal model of DFA-RIS based on sub-array, the  signal received by the $k$th user, denoted by $y_k, \forall k \in \mathcal{K}$, is derived as follows:
 \begin{small}
 	\begin{align}
 		\begin{split}
 			y_k=&\boldsymbol{\rm h}_{d,k}^H\boldsymbol{\rm x}+\boldsymbol{\rm h}_{r,k}^H\boldsymbol{\rm \Theta}_{\varrho(k)}\boldsymbol{\rm D}_{\varrho(k)}\boldsymbol{\rm A}\boldsymbol{\rm G}\boldsymbol{\rm x}\\
 			&\qquad\qquad+\boldsymbol{\rm h}_{r,k}^H\boldsymbol{\rm \Theta}_{\varrho(k)}\boldsymbol{\rm D}_{\varrho(k)}\boldsymbol{\rm A}\boldsymbol{\rm n}_{\scriptscriptstyle R\hspace{-0.2mm}I\hspace{-0.2mm}S}+n_k,\\
 			=&\boldsymbol{\rm h}_{k}^H\boldsymbol{\rm w}_ks_k+\textstyle\sum_{j\neq k,j=1}^{K}\boldsymbol{\rm h}_{k}^H\boldsymbol{\rm w}_js_j\\
 			&\qquad\qquad+\boldsymbol{\rm h}_{r,k}^H\boldsymbol{\rm \Theta}_{\varrho(k)}\boldsymbol{\rm D}_{\varrho(k)}\boldsymbol{\rm A}\boldsymbol{\rm n}_{\scriptscriptstyle R\hspace{-0.2mm}I\hspace{-0.2mm}S}+n_k,
 		\end{split}
 	\end{align}
 \end{small}
 where $\boldsymbol{\rm h}_{k}=\boldsymbol{\rm h}_{d,k}+\boldsymbol{\rm G}^H\boldsymbol{\rm A}\boldsymbol{\rm D}_{\varrho(k)}\boldsymbol{\rm A}\boldsymbol{\rm \Theta}_{\varrho(k)}^*\boldsymbol{\rm h}_{r,k}$ and
 $n_k\sim\mathcal{CN}(0,\sigma^2_k)$ represent equivalent channel and the additive complex Gaussian noise of the $k$th user, respectively.
 
 Based on the signal received by the $k$th user, we can derive the SINR of the $k$th user, as follows:
 \begin{small}
 	\begin{align}
 		\begin{split}
 			SINR_k=&\frac{\left\vert\boldsymbol{\rm h}_{k}^H\boldsymbol{\rm w}_k\right\vert^2}{\sum\limits_{j\neq k,j=1}^{K}\hspace{-1mm}\left\vert\boldsymbol{\rm h}_{k}^H\boldsymbol{\rm w}_j\right\vert^2\hspace{-2mm}+\hspace{-0.5mm}\sigma^2_{\scriptscriptstyle R\hspace{-0.2mm}I\hspace{-0.2mm}S}\left\Vert\boldsymbol{\rm A}\boldsymbol{\rm D}_{\varrho(k)}\boldsymbol{\rm h}_{r,k}^H\right\Vert^2\hspace{-0.5mm}+\hspace{-0.5mm}\sigma^2_k}
 			,
 		\end{split}
 	\end{align}
 \end{small}
 where $\Vert\boldsymbol{\rm \Theta}_{\varrho(k)}^*\boldsymbol{\rm A}\boldsymbol{\rm D}_{\varrho(k)}\boldsymbol{\rm h}_{r,k}\Vert^2=\Vert\boldsymbol{\rm A}\boldsymbol{\rm D}_{\varrho(k)}\boldsymbol{\rm h}_{r,k}\Vert^2$.
 Then, the achievable sum-rate of the multiuser MISO system can be obtained as follows:
 \begin{equation}\small
 R_{sum} = \sum_{k=1}^{K}R_k=\sum_{k=1}^{K}\mbox{log}_2(1+SINR_{k}).
 \end{equation}
 
 \section{Power Consumption Model And Problem Formulation}
 \label{sec:Power_Consumption_Problem_formulation}

 \subsection{Power Consumption Model of Sub-Array Based DFA-RIS}
 \label{subsec:Power_Consumption_Model}
 
 In terms of power consumption, we consider the power consumption at BS and DFA-RIS. To begin with, the power consumption at the BS, denoted as $P_{BS}$, includes the transmit power as well as the power consumed by the BS components, which can be represented as follows:
 \begin{small}
 	\begin{align}
 		\begin{split}
 			P_{BS}=\xi_{BS}\sum_{k=1}^{K}\left\Vert\boldsymbol{\rm w}_k\right\Vert^2+P_{BS,c},
 		\end{split}
 	\end{align}
 \end{small}
 where $\xi_{BS}$ and $P_{BS,c}$ represent the reciprocal of the energy conversion coefficient and the hardware static power of BS.
 
 As for the power consumption of DFA-RIS, it consists of the output power and the hardware static power consumption (including phase shifters, DPUs and RAs) \cite{DFA-RIS1}. To simplify the presentation, based on the definition of $\boldsymbol{\rm A}$ introduced by the sub-array architecture of DFA-RIS discussed previously, we introduce the vector $\boldsymbol{\rm A}_c=[\left\Vert\boldsymbol{\rm \alpha}_1-\boldsymbol{\rm 1}_{M_{sub}}\right\Vert_2,\cdots,\left\Vert\boldsymbol{\rm \alpha}_L-\boldsymbol{\rm 1}_{M_{sub}}\right\Vert_2]^T$, which can equivalently characterize the cardinality of $\mathcal{L}_s$ as follows:
 \begin{small}
 	\begin{align}
 		\begin{split}
 			\left\Vert\boldsymbol{\rm A}_c\right\Vert_0=\left\vert\mathcal{L}_s\right\vert,
 		\end{split}
 	\end{align}
 \end{small}
 where $\ell_0$-norm represents the number of non-zero elements in $\boldsymbol{\rm A}_c$. Therefore, the power consumption at the DFA-RIS based on sub-array can be written as follows:
 \begin{small}
 	\begin{align}
 		\begin{split}
 			P_{RIS}=&\xi_{RIS}\left(\sum_{k=1}^{K}\left\Vert\boldsymbol{\rm A}\boldsymbol{\rm G}\boldsymbol{\rm w}_k\right\Vert^2+\sigma^2_{\scriptscriptstyle R\hspace{-0.2mm}I\hspace{-0.2mm}S}\left\Vert\boldsymbol{\rm A}\right\Vert^2_F\right)\\
 			&\quad\qquad+2MP_{PS}+MP_D+\left\Vert \boldsymbol{\rm A}_c\right\Vert_0M_{RIS}P_{A},
 		\end{split}
 	\end{align}
 \end{small}
 where $P_{PS}$, $P_D$, $P_{A}$, and $\xi_{RIS}$  denote the hardware static power of the phase shifter, PDU, and RA, and the reciprocal of the energy conversion efficiency at DFA-RIS, respectively. As a consequence, the total power consumption of the system, denoted by $P_{tot}$,  can be expressed as $P_{tot}=P_{BS}+P_{RIS}$. 
 Further, the energy efficiency (EE) of the system, denoted by $\eta_{EE}$, can be described as follows:
 \begin{align}
 \begin{split}
 \eta_{EE} = \frac{R_{sum}}{P_{tot}}.
 \end{split}
 \end{align}
 \subsection{Problem Formulation}
 \label{sec:Problem_formulation_1}
 
 In this subsection, we formulate the optimization problem to maximize the EE of the system by jointly designing the beamforming of BS, the amplifying coefficient, the power splitting coefficient, and the phase-shifters of DFA-RIS, subject to maximum transmit power, maximum power consumption at DFA-RIS, maximum amplification amplitude, and minimum individual quality-of-service (QoS). In addition, we investigate the case of channel conditions for perfect CSI and imperfect CSI, respectively.
 
 \subsubsection{The Case of Perfect CSI} 

  To begin with, we investigate the case where the CSI of all channels is perfectly known. The optimization results obtained in such an ideal case can be utilized as performance upper bounds for reference analysis. For convenience, we define $\boldsymbol{\rm w}=[\boldsymbol{\rm w}_1^T,\cdots,\boldsymbol{\rm w}_K^T]^T$ and $\boldsymbol{\rm \alpha}=[\boldsymbol{\rm \alpha}_1^T,\cdots,\boldsymbol{\rm \alpha}_L^T]^T$. Accordingly, we can formulate the maximization problem $\textbf{\textit{P}1}$ as follows:
  \begin{subequations}\label{Problem:perfect1}
  	\begin{alignat}{2}
  	\textbf{\textit{P}1:}
  	&\mathop{\max}\limits_{\boldsymbol{\rm {w}},\boldsymbol{\rm {\theta}}_R,\boldsymbol{\rm {\theta}}_T,\boldsymbol{\rm {\alpha}},\boldsymbol{\rm {\beta}}} f_P(\boldsymbol{\rm {w}},\boldsymbol{\rm {\theta}}_R,\boldsymbol{\rm {\theta}}_T,\boldsymbol{\rm {\alpha}},\boldsymbol{\rm {\beta}}) = \eta_{EE}  \notag \\
  	{\rm{s.t.}}:&1).\ \xi_{BS}\sum_{k=1}^{K}\left\Vert\boldsymbol{\rm {w}}_k\right\Vert^2\le P_{\max}^{BS};\\
  	&2).\ \xi_{RIS}\left(\sum_{k=1}^{K}\left\Vert\boldsymbol{\rm A}\boldsymbol{\rm G}\boldsymbol{\rm w}_k\right\Vert^2+\sigma^2_{\scriptscriptstyle R\hspace{-0.2mm}I\hspace{-0.2mm}S}\left\Vert\boldsymbol{\rm A}\right\Vert^2_F\right)+2MP_{PS}\nonumber\\
  	&\qquad+MP_D+\left\Vert \boldsymbol{\rm A}_c\right\Vert_0M_{RIS}P_{A}\le P_{\max}^{RIS}; \\
  	&3).\ 0\le\alpha_m\le\alpha_{\max}, \forall m \in \mathcal{M};\\
  	&4).\  0\le\beta_m\le 1, \forall m \in \mathcal{M};\\
  	&5).\  \left\vert\theta_{R,m}\right\vert = 1, \left\vert\theta_{T,m}\right\vert = 1, \forall m \in \mathcal{M};\\
  	&6).\  R_k\ge R_{\min}, \forall k \in \mathcal{K},
  	\end{alignat}
  \end{subequations}
 where $P_{\max}^{BS}$, $P_{\max}^{RIS}$, and $R_{\min}$ indicate the maximum transmit power, maximum power consumption at DFA-RIS, and minimum QoS (rate) requirement of the $k$th user. 
 
 Note that the optimization problem $\textbf{\textit{P}1}$ is a nonconvex problem since the optimization variables are coupled in the objective function of $\textbf{\textit{P}1}$. Moreover, the optimization problem is generally difficult to solve due to the presence of fractional forms in the objective function as well as the constraint (\ref{Problem:perfect1}{f}).
 
 \subsubsection{The Case of Imperfect CSI}
 
 In practice, RIS lacks the necessary signal processing capabilities, which limits the individual estimation of the channels connected to it. Specifically, only the CSI of the combined channel $\boldsymbol{\rm h}_{k}$ needs to be known at the BS for optimizing $\boldsymbol{\rm w}$. Therefore, it is still reasonable to assume that the perfect CSI of $\boldsymbol{\rm h}_{k}$ is available at the BS. However, in the DFA-RIS assisted system, the optimization of the variables at DFA-RIS depends on the individual CSIs of $\boldsymbol{\rm h}_{d,k}$, $\boldsymbol{\rm G}$, and $\boldsymbol{\rm h}_{r,k}$, respectively, which are difficult to estimate with high precision. To address this issue, we assume the imperfect CSI of channels, where the estimated channels $\boldsymbol{\rm \widehat{h}}_{d,k}$, $\boldsymbol{\rm \widehat{G}}$, and $\boldsymbol{\rm \widehat{h}}_{r,k}$
 are expressed as follows:
 \begin{align}
 \begin{split}
 \boldsymbol{\rm h}_{d,k} \hspace{-1mm}= \hspace{-1mm}\boldsymbol{\rm \widehat{h}}_{d,k}\hspace{-1mm}+\hspace{-1mm}\boldsymbol{\rm \Delta h}_{d,k},
 \boldsymbol{\rm G} \hspace{-1mm}=\hspace{-1mm} \boldsymbol{\rm \widehat{G}}\hspace{-1mm}+\hspace{-1mm}\boldsymbol{\rm \Delta G},
 \boldsymbol{\rm h}_{r,k}\hspace{-1mm} =\hspace{-1mm} \boldsymbol{\rm \widehat{h}}_{r,k}\hspace{-1mm}+\hspace{-1mm}\boldsymbol{\rm \Delta h}_{r,k},
 \end{split}
 \end{align}
 where $\boldsymbol{\rm \Delta h}_{d,k}$, $\boldsymbol{\rm \Delta G}$, and $\boldsymbol{\rm \Delta h}_{r,k}$ denote the corresponding uncertain estimation errors. Based on the above channel model, the actual channel coefficients $\boldsymbol{\rm h}_{d,k}$, $\boldsymbol{\rm G}$ and $\boldsymbol{\rm h}_{r,k}$ can be constructed as realizations of the sample space $\mathcal{S}=\{\boldsymbol{\rm h}_{d,k}(\vartheta), \boldsymbol{\rm G}(\vartheta), \boldsymbol{\rm h}_{r,k}(\vartheta), \forall k, \forall \vartheta\}$, where $\vartheta$ denotes the index of random realizations drawn from $\mathcal{S}$. It can be seen that the actual channels depend on their corresponding estimates and estimation errors. 
 
 
 Based on the above imperfect CSI settings of the channels, we can formulate the maximization problem $\textbf{\textit{P}2}$ as follows:
 \begin{subequations}\label{Problem:Imperfect2}
 	\begin{alignat}{2}
 	\textbf{\textit{P}2:}
 	&\mathop{\max}\limits_{\boldsymbol{\rm {\theta}}_R,\boldsymbol{\rm {\theta}}_T,\boldsymbol{\rm {\alpha}},\boldsymbol{\rm {\beta}}}f_I(\boldsymbol{\rm {\theta}}_R,\boldsymbol{\rm {\theta}}_T,\boldsymbol{\rm {\alpha}},\boldsymbol{\rm {\beta}})=\mathbb{E}_{\vartheta}\left[\mathop{\max}\limits_{\boldsymbol{\rm {w}}}f_P(\boldsymbol{\rm {w}};\vartheta)\right]   \notag \\
 	{\rm{s.t.}}:&1).\ (\ref{Problem:perfect1}{\text{a}}),(\ref{Problem:perfect1}{\text{c}}), (\ref{Problem:perfect1}{\text{d}}), \text{and}\ (\ref{Problem:perfect1}{\text{e}}), \notag\\
 	&2).\ \mathbb{E}_{\vartheta}\left[f_{c1}(\boldsymbol{\rm {w}},\boldsymbol{\rm {\alpha}};\vartheta)\right]\le P_{\max}^{RIS}; \\
 	&3).\  \mathbb{E}_{\vartheta}\left[R_k(\boldsymbol{\rm {w}},\boldsymbol{\rm {\theta}}_R,\boldsymbol{\rm {\theta}}_T,\boldsymbol{\rm {\alpha}},\boldsymbol{\rm {\beta}};\vartheta)\right]\hspace{-1mm}\ge R_{\min}, \forall k \in \mathcal{K},
 	\end{alignat}
 \end{subequations}
 where  $f_{c1}(\boldsymbol{\rm {w}},\boldsymbol{\rm {\alpha}};\vartheta)$ and $R_k(\boldsymbol{\rm {w}},\boldsymbol{\rm {\theta}}_R,\boldsymbol{\rm {\theta}}_T,\boldsymbol{\rm {\alpha}},\boldsymbol{\rm {\beta}};\vartheta)$ indicate the left-hand sides (LHS) of constraints (\ref{Problem:perfect1}{b}) and (\ref{Problem:perfect1}{f}) at the $\vartheta$-th random channel realization, respectively. It can be seen that, on the basis of the optimization problem $\textbf{\textit{P}1}$ in the perfect CSI case, $\textbf{\textit{P}2}$ is constructed as a stochastic optimization problem by further introducing expectations. Thus, the optimization problem is generally intractable.
 
 \section{PDD-Based Alternating Optimization In The Case Of Perfect CSI}
 \label{sec:Alternating_optimization_perfect_CSI}
 
 The objective function of $\textbf{\textit{P}1}$ is a typical fractional function which can be equivalently transformed into a more solvable form by the FP algorithm. Then the transformed problem is divided into a number of subproblems, with each optimization variable being designed alternatively.
 \subsection{Fractional Programming Transform}
 \label{subsec:FP}
 
 The fractional objective function is first converted to an equivalent form by applying the Dinkelbach  algorithm \cite{Dinkelbach}. The optimal EE $\eta_{EE}^{\star}$ satisfies
 \begin{align}
 \mathop{\max}\limits_{\boldsymbol{\rm {w}},\boldsymbol{\rm {\theta}}_R,\boldsymbol{\rm {\theta}}_T,\boldsymbol{\rm {\alpha}},\boldsymbol{\rm {\beta}}} \left(R_{sum}-\eta_{EE}^{\star}P_{tot}\right)=0.
 \end{align}
 Thus, $\textbf{\textit{P}1}$ can be equivalently transformed as follows:
 \begin{subequations}\label{Problem:perfect2}\small
 	\begin{alignat}{2}
 	\mathop{\max}\limits_{\boldsymbol{\rm {w}},\boldsymbol{\rm {\theta}}_R,\boldsymbol{\rm {\theta}}_T,\boldsymbol{\rm {\alpha}},\boldsymbol{\rm {\beta}}} &f_P(\boldsymbol{\rm {w}},\boldsymbol{\rm {\theta}}_R,\boldsymbol{\rm {\theta}}_T,\boldsymbol{\rm {\alpha}},\boldsymbol{\rm {\beta}}) = R_{sum}-\eta_{EE}P_{tot}  \notag \\
 	{\rm{s.t.}}:&(\ref{Problem:perfect1}{\text{a}}),(\ref{Problem:perfect1}{\text{b}}), (\ref{Problem:perfect1}{\text{c}}), (\ref{Problem:perfect1}{\text{d}}), (\ref{Problem:perfect1}{\text{e}}), \text{and}\ (\ref{Problem:perfect1}{\text{f}}),\nonumber
 	\end{alignat}
 \end{subequations}
 where $\eta_{EE}^{\star}$ can be solved by alternating optimization.
 
 However the objective function remains in the form of a fraction which is intractable to solve. Further, based on the Lagrangian Dual Transform and Quadratic Transform \cite{Quadratic_Transform}, we introduce the auxiliary variables $\boldsymbol{\rm {\gamma}}=[\gamma_1,\cdots,\gamma_K]^T$ and $\boldsymbol{\rm {\nu}}=[\nu_1,\cdots,\nu_K]^T$ to equivalently transform the objective function of $\textbf{\textit{P}1}$ as follows:
 \begin{small}
 	\begin{align}
 		\begin{split}
 			f_P^I(\boldsymbol{\rm {w}},\boldsymbol{\rm {\gamma}},\boldsymbol{\rm {\nu}},\boldsymbol{\rm {\theta}}_R,\boldsymbol{\rm {\theta}}_T,\boldsymbol{\rm {\alpha}},\boldsymbol{\rm {\beta}}) =&g_P(\boldsymbol{\rm {w}},\boldsymbol{\rm {\gamma}},\boldsymbol{\rm {\nu}},\boldsymbol{\rm {\theta}}_R,\boldsymbol{\rm {\theta}}_T,\boldsymbol{\rm {\alpha}},\boldsymbol{\rm {\beta}})\\
 			&-\eta_{EE}P_{tot},
 		\end{split}
 	\end{align}
 \end{small}
 where 
 \begin{small}
 	\begin{align}
 		\begin{split}
 			&g_P(\boldsymbol{\rm {w}},\boldsymbol{\rm {\gamma}},\boldsymbol{\rm {\nu}},\boldsymbol{\rm {\theta}}_R,\boldsymbol{\rm {\theta}}_T,\boldsymbol{\rm {\alpha}},\boldsymbol{\rm {\beta}}) \\
 			&=\sum_{k=1}^{K}\left[\mbox{ln}(1+\gamma_k)-\gamma_k+2\sqrt{1+\gamma_k}\mbox{Re}\left\{\nu_k^*\boldsymbol{\rm {h}}_k^H\boldsymbol{\rm {w}}_k\right\}\right.\\
 			&\quad-\hspace{-1mm}\left.\left\vert\nu_k\right\vert^2\hspace{-1mm}\left(\hspace{-1mm}\textstyle\sum_{j=1}^{K}\left\vert\boldsymbol{\rm {h}}_k^H\boldsymbol{\rm {w}}_j\right\vert^2\hspace{-1mm}+\hspace{-0.5mm}\sigma^2_{\scriptscriptstyle R\hspace{-0.2mm}I\hspace{-0.2mm}S}\Vert\boldsymbol{\rm A}\boldsymbol{\rm D}_{\varrho(k)}\boldsymbol{\rm h}_{r,k}\Vert^2\hspace{-1mm}+\hspace{-1mm}\sigma^2_{k}\hspace{-1mm}\right)\hspace{-1mm}\right]\hspace{-1mm},
 		\end{split}
 	\end{align}
 \end{small}
 which is an equivalent replacement of the $R_{sum}$ part of the objective function. Then similarly, the fractional constraint (\ref{Problem:perfect1}{f}) is equivalently replaced as follows:
 \begin{small}
 	\begin{align}
 		\begin{split}\label{constraint:12f}
 			g_{P,k}(\boldsymbol{\rm {w}},\boldsymbol{\rm {\gamma}},\boldsymbol{\rm {\nu}},\boldsymbol{\rm {\theta}}_R,\boldsymbol{\rm {\theta}}_T,\boldsymbol{\rm {\alpha}},\boldsymbol{\rm {\beta}})\ge R_{\min}, \forall k\in \mathcal{K},
 		\end{split}
 	\end{align}
 \end{small}
 where
 \begin{small}
 	\begin{align}
 		\begin{split}
 			&g_{P,k}(\boldsymbol{\rm {w}},\boldsymbol{\rm {\gamma}},\boldsymbol{\rm {\nu}},\boldsymbol{\rm {\theta}}_R,\boldsymbol{\rm {\theta}}_T,\boldsymbol{\rm {\alpha}},\boldsymbol{\rm {\beta}}) \\
 			&=\mbox{ln}(1+\gamma_k)-\gamma_k+2\sqrt{1+\gamma_k}\mbox{Re}\left\{\nu_k^*\boldsymbol{\rm {h}}_k^H\boldsymbol{\rm {w}}_k\right\}\\
 			&\quad-\left\vert\nu_k\right\vert^2\hspace{-1mm}\left(\textstyle\sum_{j=1}^{K}\left\vert\boldsymbol{\rm {h}}_k^H\boldsymbol{\rm {w}}_j\right\vert^2\hspace{-1.5mm}+\hspace{-1mm}\sigma^2_{\scriptscriptstyle R\hspace{-0.2mm}I\hspace{-0.2mm}S}\Vert\boldsymbol{\rm A}\boldsymbol{\rm D}_{\varrho(k)}\boldsymbol{\rm h}_{r,k}\Vert^2\hspace{-1mm}+\hspace{-1mm}\sigma^2_{k}\hspace{-1mm}\right)\hspace{-1mm}.
 		\end{split}
 	\end{align}
 \end{small}

 With the derivation mentioned above, the optimization problem $\textbf{\textit{P}1}$ is rewritten as follows:
 \begin{subequations}\label{Problem:perfectA}
 	\begin{alignat}{2}
 	\textbf{\textit{P}1-(A):}&\mathop{\max}\limits_{\boldsymbol{\rm {w}},\boldsymbol{\rm {\gamma}},\boldsymbol{\rm {\nu}},\boldsymbol{\rm {\theta}}_R,\boldsymbol{\rm {\theta}}_T,\boldsymbol{\rm {\alpha}},\boldsymbol{\rm {\beta}}} f_P^I(\boldsymbol{\rm {w}},\boldsymbol{\rm {\gamma}},\boldsymbol{\rm {\nu}},\boldsymbol{\rm {\theta}}_R,\boldsymbol{\rm {\theta}}_T,\boldsymbol{\rm {\alpha}},\boldsymbol{\rm {\beta}})   \notag \\
 	{\rm{s.t.}}:&(\ref{Problem:perfect1}{\text{a}}),(\ref{Problem:perfect1}{\text{b}}), (\ref{Problem:perfect1}{\text{c}}), (\ref{Problem:perfect1}{\text{d}}), (\ref{Problem:perfect1}{\text{e}}), \text{and}\ (\ref{constraint:12f}).\nonumber
 	\end{alignat}
 \end{subequations}
 
 To deal with the problem $\textbf{\textit{P}1-(A)}$ efficiently, we propose an effective AO algorithm to update the optimization variables alternately, which is derived in details as follows.
 \subsection{Update $\boldsymbol{\rm {w}}$, $\boldsymbol{\rm {\gamma}}$, And $\boldsymbol{\rm {\nu}}$}
 \label{subsec:w_g_v}
 
 \subsubsection{Update auxiliary variables $\boldsymbol{\rm {\gamma}}$ and $\boldsymbol{\rm {\nu}}$}
 
 $\textbf{\textit{P}1-(A)}$ demonstrates block-wise convexity, with the optimization of both the optimization variables and the introduced auxiliary variables being executed within the framework of AO. Therefore, with the other variables fixed, the optimal $\gamma_k^{\star}$ and $\nu_k^{\star}$ can be obtained in a closed-form by setting $\partial f_P^I/\partial \gamma_k$ and $\partial f_P^I/\partial\nu_k$ to zero, respectively, as follows:
 \begin{small}
 	\begin{align}\label{gamma_k}
 		\gamma_k^{\star}=&\frac{\left\vert\boldsymbol{\rm h}_{k}^H\boldsymbol{\rm w}_k\right\vert^2}{\sum\limits_{j\neq k,j=1}^{K}\hspace{-1mm}\left\vert\boldsymbol{\rm h}_{k}^H\boldsymbol{\rm w}_j\right\vert^2\hspace{-2mm}+\hspace{-0.5mm}\sigma^2_{\scriptscriptstyle R\hspace{-0.2mm}I\hspace{-0.2mm}S}\left\Vert\boldsymbol{\rm A}\boldsymbol{\rm D}_{\varrho(k)}\boldsymbol{\rm h}_{r,k}^H\right\Vert^2\hspace{-0.5mm}+\hspace{-0.5mm}\sigma^2_k}
 		,\\
 		\nu_k^{\star}=&\frac{\sqrt{1-\gamma_k}\boldsymbol{\rm h}_{k}^H\boldsymbol{\rm w}_k}{\sum\limits_{j=1}^{K}\hspace{-1mm}\left\vert\boldsymbol{\rm h}_{k}^H\boldsymbol{\rm w}_j\right\vert^2\hspace{-2mm}+\hspace{-0.5mm}\sigma^2_{\scriptscriptstyle R\hspace{-0.2mm}I\hspace{-0.2mm}S}\left\Vert\boldsymbol{\rm A}\boldsymbol{\rm D}_{\varrho(k)}\boldsymbol{\rm h}_{r,k}^H\right\Vert^2\hspace{-0.5mm}+\hspace{-0.5mm}\sigma^2_k}.\label{nu_k}
 	\end{align}
 \end{small}

 \subsubsection{Update transmit beamforming $\boldsymbol{\rm {w}}$}
 
 When the other variables are fixed, the subproblem to optimize $\boldsymbol{\rm {w}}$ is described as follows:
 \begin{subequations}\label{Problem:perfect1B1}
 	\begin{alignat}{2}
 	\mathop{\max}\limits_{\boldsymbol{\rm {w}}}& f_P^I(\boldsymbol{\rm {w}};\boldsymbol{\rm {\gamma}},\boldsymbol{\rm {\nu}},\boldsymbol{\rm {\theta}}_R,\boldsymbol{\rm {\theta}}_T,\boldsymbol{\rm {\alpha}},\boldsymbol{\rm {\beta}})   \notag \\
 	{\rm{s.t.}}:&\  (\ref{Problem:perfect1}{\text{a}}),(\ref{Problem:perfect1}{\text{b}}), \text{and}\ (\ref{constraint:12f}).\nonumber
 	\end{alignat}
 \end{subequations}
 For brevity, by omitting constant terms independent of $\boldsymbol{\rm {w}}$, the subproblem is further rewritten as follows:
 \begin{subequations}\label{Problem:perfectw}\small
 	\begin{alignat}{2}
 	\textbf{\textit{P}1-(B):}
 	&\mathop{\max}\limits_{\boldsymbol{\rm {w}}} \ \mbox{Re}\left\{\boldsymbol{\rm {y}}^H\boldsymbol{\rm {w}}\right\}-\boldsymbol{\rm {w}}^H\boldsymbol{\rm {Y}}\boldsymbol{\rm {w}} \notag \\
 	{\rm{s.t.}}:&1).\ \boldsymbol{\rm {w}}^H\boldsymbol{\rm {w}}\le \widetilde{P}_{\max}^{BS};\\
 	&2).\ \boldsymbol{\rm {w}}^H\boldsymbol{\rm {Z}}\boldsymbol{\rm {w}}\le \widetilde{P}_{\max}^{RIS}; \\
 	&3).\  2\sqrt{1+\gamma_k}\mbox{Re}\left\{\nu_k^*\boldsymbol{\rm {h}}_k^H\boldsymbol{\rm {w}}_k\right\}
 	\hspace{-1mm}-\hspace{-1mm}\left\vert\nu_k\right\vert^2\left\vert\boldsymbol{\rm {h}}_k^H\boldsymbol{\rm {w}}_k\right\vert^2\hspace{-1mm}\ge c_{0,k},\nonumber\\
 	 &\quad\forall k \in \mathcal{K},
 	\end{alignat}
 \end{subequations}
 where we define 
 \begin{small}
 	\begin{align}
 	&\boldsymbol{\rm {y}}\triangleq[\boldsymbol{\rm {y}}_1^T,\cdots,\boldsymbol{\rm {y}}_K^T]^T,\boldsymbol{\rm {y}}_k=2\sqrt{1+\gamma_k}\nu_k\boldsymbol{\rm {h}}_k,\notag\\
 	&\boldsymbol{\rm {Y}}\triangleq\boldsymbol{\rm {I}}_k\hspace{-1mm}\otimes\hspace{-1mm}\left(\sum_{k=1}^{K}\vert \nu_k\vert^2\boldsymbol{\rm {h}}_k\boldsymbol{\rm {h}}_k^H\hspace{-1mm}+\hspace{-1mm}\eta_{EE}\xi_{BS}\boldsymbol{\rm {I}}_N\hspace{-1mm}+\hspace{-1mm}\eta_{EE}\xi_{BS}\boldsymbol{\rm {G}}^H\hspace{-1mm}\boldsymbol{\rm {A}}\boldsymbol{\rm {A}}\boldsymbol{\rm {G}}\hspace{-1mm}\right)\hspace{-1mm},\notag\\
 	&\boldsymbol{\rm {Z}}\triangleq\boldsymbol{\rm {I}}_k\otimes\left(\boldsymbol{\rm {G}}^H\boldsymbol{\rm {A}}\boldsymbol{\rm {A}}\boldsymbol{\rm {G}}\right),~ \widetilde{P}_{\max}^{BS} \triangleq \xi_{BS}^{-1} P_{\max}^{BS},\notag\\
 	&\widetilde{P}_{\max}^{RIS} \triangleq\xi_{RIS}^{-1}\left(P_{\max}^{RIS}-2MP_{PS}-MP_{D}-\xi_{RIS}\sigma^2_{\scriptscriptstyle R\hspace{-0.2mm}I\hspace{-0.2mm}S}\left\Vert\boldsymbol{\rm A}\right\Vert^2_F\right.\notag\\
 	&\qquad\qquad-\left\Vert \boldsymbol{\rm A}_c\right\Vert_0M_{RIS}P_{A}\Big),\notag\\
 	&c_{0,k}\triangleq R_{\min}-\mbox{ln}(1+\gamma_k)+\gamma_k+\left\vert\nu_k\right\vert^2\textstyle\sum_{j\neq k, j=1}^{K}\left\vert\boldsymbol{\rm {h}}_k^H\boldsymbol{\rm {w}}_j\right\vert^2\notag\\
 	&\qquad\qquad+\sigma^2_{\scriptscriptstyle R\hspace{-0.2mm}I\hspace{-0.2mm}S}\left\vert\nu_k\right\vert^2\Vert\boldsymbol{\rm A}\boldsymbol{\rm D}_{\varrho(k)}\boldsymbol{\rm h}_{r,k}\Vert^2+\left\vert\nu_k\right\vert^2\sigma^2_{k}.
 	\end{align}
 \end{small}
 $\textbf{\textit{P}1-(B)}$ is a a standard quadratic constraint quadratic programming (QCQP) problem, thus we can use convex program solvers such as CVX \cite{CVX-intro} to obtain the optimal solutions.
 
 \subsection{Update $\boldsymbol{\rm {\theta}}_R$, $\boldsymbol{\rm {\theta}}_T$, $\boldsymbol{\rm {\alpha}}$, And $\boldsymbol{\rm {\beta}}$}
 \label{subsec:a_b_p}
 
 \subsubsection{Update the power splitting parameter $\boldsymbol{\rm {\beta}}$}
 \label{subsubsec:beta}
 
 With other variables being given, we further investigate the optimization of the power splitting parameter $\boldsymbol{\rm {\beta}}$. To begin with, we introduce the new definitions $\kappa_{k,j}\triangleq\boldsymbol{\rm {h}}_{d,k}^H\boldsymbol{\rm {w}}_j$, $\boldsymbol{\rm {\Upsilon}}_{k,j}\triangleq\mbox{Diag}(\boldsymbol{\rm {h}}_{r,k}^*)\boldsymbol{\rm {\Theta}}_{\varrho(k)}\boldsymbol{\rm {A}}\boldsymbol{\rm {G}}\boldsymbol{\rm {w}}_j$,  $\boldsymbol{\rm {d}}_{\varrho(k)}\triangleq\boldsymbol{\rm {\beta}}$ if $k\in\mathcal{K_R}$, and $\boldsymbol{\rm {d}}_{\varrho(k)}\triangleq\sqrt{1-\boldsymbol{\rm {\beta}}^2}$ if $k\in\mathcal{K_T}$. Then, by dropping the irrelevant terms with respect to $\boldsymbol{\rm {\beta}}$, the terms of the objective function $f_P^I(\boldsymbol{\rm {\beta}}; \boldsymbol{\rm {w}},\boldsymbol{\rm {\gamma}},\boldsymbol{\rm {\nu}},\boldsymbol{\rm {\theta}}_R,\boldsymbol{\rm {\theta}}_T,\boldsymbol{\rm {\alpha}})$ are rewritten as follows:
 \begin{small}
 	\begin{align}\label{Eqs:fB1}
 		&-\hspace{-2mm}\sum_{k=1}^{K}\hspace{-1mm}\left\vert\nu_k\right\vert^2\hspace{-1mm}\sum_{j=1}^{K}\hspace{-1mm}\left\vert\boldsymbol{\rm {h}}_k^H\boldsymbol{\rm {w}}_j\right\vert^2\hspace{-2mm}=\hspace{-1mm}-\hspace{-4mm}\sum_{\varrho\in\{R,T\}}\hspace{-3mm}\left(\boldsymbol{\rm {d}}_{\varrho}^H\boldsymbol{\rm {\bar{U}}}_{\varrho}\boldsymbol{\rm {d}}_{\varrho}\hspace{-1mm}+\hspace{-1mm}2\mbox{Re}\left\{\boldsymbol{\rm {\bar{u}}}_{\varrho}^H\boldsymbol{\rm {d}}_{\varrho}\right\}\right)\hspace{-1mm}+\hspace{-1mm}c_1,\\\label{Eqs:fB2}
 		&\sum_{k=1}^{K}2\sqrt{1+\gamma_k}\mbox{Re}\left\{\nu_k\boldsymbol{\rm {h}}_k^H\boldsymbol{\rm {w}}_j\right\}=2\hspace{-4mm}\sum_{\varrho\in\{R,T\}}\hspace{-3mm}\mbox{Re}\hspace{-1mm}\left\{\boldsymbol{\rm {\widetilde{u}}}_{\varrho}^H\boldsymbol{\rm {d}}_{\varrho}\right\}\hspace{-1mm}+\hspace{-1mm}c_2,\\\label{Eqs:fB3}
 		&\sum_{k=1}^{K}\left\vert\nu_k\right\vert^2\sigma^2_{\scriptscriptstyle R\hspace{-0.2mm}I\hspace{-0.2mm}S}\Vert\boldsymbol{\rm A}\boldsymbol{\rm D}_{\varrho(k)}\boldsymbol{\rm h}_{r,k}\Vert^2=\hspace{-4mm}\sum_{\varrho\in\{R,T\}}\hspace{-3mm}\boldsymbol{\rm {d}}_{\varrho}^H\boldsymbol{\rm {\widetilde{U}}}_{\varrho}\boldsymbol{\rm {d}}_{\varrho},
 	\end{align}
 \end{small}
 where the parameters are defined as follows ($\varrho\in\left\{R,T\right\}$):
 \begin{small}
 	 \begin{align}
 	&\boldsymbol{\rm {\bar{U}}}_{\varrho}\hspace{-1mm}\triangleq\hspace{-1mm}\sum_{k\in K_\varrho}\left\vert\nu_k\right\vert^2\sum_{j=1}^{K}\boldsymbol{\rm {\Upsilon}}_{k,j}\boldsymbol{\rm {\Upsilon}}_{k,j}^H,~\boldsymbol{\rm {\bar{u}}}_{\varrho}\triangleq\hspace{-1mm}\sum_{k\in K_\varrho}\hspace{-1mm}\left\vert\nu_k\right\vert^2\sum_{j=1}^{K}\kappa_{k,j}\boldsymbol{\rm {\Upsilon}}_{k,j},\notag\\
 	&\boldsymbol{\rm {\widetilde{U}}}_{\varrho}\triangleq\hspace{-1mm}\sum_{k\in K_\varrho}\sigma^2_{\scriptscriptstyle R\hspace{-0.2mm}I\hspace{-0.2mm}S}\mbox{Diag}\left(\vert\boldsymbol{\rm {A}}\boldsymbol{\rm {h}}_{r,k}\vert^2\right),\notag\\
 	&\boldsymbol{\rm {\widetilde{u}}}_{\varrho}\hspace{-0.5mm}\triangleq\hspace{-2mm}\sum_{k\in K_\varrho}\hspace{-3mm}\sqrt{1\hspace{-0.5mm}+\hspace{-0.5mm}\gamma_k}\nu_k^*\mbox{Diag}(\boldsymbol{\rm {h}}_{r,k}^*)\boldsymbol{\rm {\Theta}}_{\varrho(k)}\boldsymbol{\rm {A}}\boldsymbol{\rm {G}}\boldsymbol{\rm {w}}_k,\notag\\
 	&c_1 \hspace{-1mm}\triangleq\hspace{-1mm}-\hspace{-1mm}\sum_{k=1}^{K}\hspace{-1mm}\left\vert\nu_k\right\vert^2\sum_{j=1}^{K}\left\vert\kappa_{k,j}\right\vert^2\hspace{-1mm},c_2 \triangleq 2\hspace{-1mm}\sum_{k=1}^{K}\hspace{-1mm}\sqrt{1\hspace{-0.5mm}+\hspace{-0.5mm}\gamma_k}\mbox{Re}\left\{\nu_k^*\boldsymbol{\rm {h}}_{d,k}^H\boldsymbol{\rm {w}}_{k}\right\}\hspace{-1mm},
 	\end{align}
 \end{small}
 Substituting Eq.~(\ref{Eqs:fB1}), Eq.~(\ref{Eqs:fB2}) and Eq.~(\ref{Eqs:fB3}) into the objective function, we can reconstruct the optimization subproblem as follows:
 \begin{subequations}\label{Problem:perfectB}\small
 	\begin{alignat}{2}
 	\mathop{\max}\limits_{\boldsymbol{\rm {\beta}}}& \ \sum_{\varrho\in\{R,T\}}\left(\boldsymbol{\rm {d}}_{\varrho}^H\boldsymbol{\rm {U}}_{\varrho}\boldsymbol{\rm {d}}_{\varrho}+2\mbox{Re}\left\{\boldsymbol{\rm {u}}_{\varrho}^H\boldsymbol{\rm {d}}_{\varrho}\right\}\right) \notag \\
 	{\rm{s.t.}}:&1).\ \  0\le\beta_m\le 1, \forall m \in \mathcal{M};\\
 	&2).\  R_k\ge R_{\min}, \forall k \in \mathcal{K},
 	\end{alignat}
 \end{subequations}
 where $\boldsymbol{\rm {U}}_{\varrho}=\boldsymbol{\rm {\bar{U}}}_{\varrho}+\boldsymbol{\rm {\widetilde{U}}}_{\varrho}$ and $\boldsymbol{\rm {u}}_{\varrho}=\boldsymbol{\rm {\bar{u}}}_{\varrho}-\boldsymbol{\rm {\widetilde{u}}}_{\varrho}$. 
 
 However, the subproblem remains intractable due to the constraint (\ref{Problem:perfectB}{b}). Hence, adopting a similar treatment for the objective function, we introduce the definition as follows:
 \begin{small}
 	\begin{align}
 	&\boldsymbol{\rm {U}}_{k}\triangleq\left\vert\nu_k\right\vert^2\hspace{-0.5mm}\left(\hspace{-0.5mm}\sum_{j=1}^{K}\boldsymbol{\rm {\Upsilon}}_{k,j}\boldsymbol{\rm {\Upsilon}}_{k,j}^H+\sigma^2_{\scriptscriptstyle R\hspace{-0.2mm}I\hspace{-0.2mm}S}\mbox{Diag}\left(\vert\boldsymbol{\rm {A}}\boldsymbol{\rm {h}}_{r,k}\vert^2\right)\hspace{-1mm}\right)\hspace{-1mm},\notag\\
 	&\boldsymbol{\rm {u}}_{k}\hspace{-1mm}\triangleq\hspace{-1mm}\left\vert\nu_k\right\vert^2\sum_{j=1}^{K}\hspace{-1mm}\kappa_{k,j}\hspace{-0.5mm}\boldsymbol{\rm {\Upsilon}}_{k,j}\hspace{-1mm}-\hspace{-1mm}\sqrt{1\hspace{-0.5mm}+\hspace{-0.5mm}\gamma_k}\nu_k^*\mbox{Diag}(\hspace{-0.5mm}\boldsymbol{\rm {h}}_{r,k}^*\hspace{-0.5mm})\boldsymbol{\rm {\Theta}}_{\varrho(k)}\boldsymbol{\rm {A}}\boldsymbol{\rm {G}}\boldsymbol{\rm {w}}_k,\notag\\
 	&c_{1,k} \hspace{-1mm}\triangleq\hspace{-0.5mm}-\hspace{-1mm}\left\vert\nu_k\right\vert^2\hspace{-0.5mm}\sum_{j=1}^{K}\hspace{-0.5mm}\left\vert\kappa_{k,j}\right\vert^2\hspace{-1mm},~c_{2,k} \hspace{-1mm}\triangleq\hspace{-0.5mm} 2\hspace{-0.5mm}\sqrt{1\hspace{-0.5mm}+\hspace{-0.5mm}\gamma_k}\mbox{Re}\hspace{-0.5mm}\left\{\hspace{-0.5mm}\nu_k^*\boldsymbol{\rm {h}}_{d,k}^H\boldsymbol{\rm {w}}_{k}\hspace{-0.5mm}\right\}\hspace{-1mm}.
 	\end{align}
 \end{small}
 Moreover, we replace constraint (\ref{Problem:perfectB}{b}) with the following constraint:
 \begin{align}\label{Eqs:constraintf}
 \boldsymbol{\rm {d}}_{\varrho}^H\boldsymbol{\rm {U}}_{k}\boldsymbol{\rm {d}}_{\varrho}+2\mbox{Re}\left\{\boldsymbol{\rm {u}}_{k}^H\boldsymbol{\rm {d}}_{\varrho}\right\}\le c_{3,k},~\forall k \in \mathcal{K},
 \end{align}
 where $c_{3,k}=-R_{\min}+\mbox{ln}(1+\gamma_k)-\gamma_k+c_{1,k}+c_{2,k}-\sigma_k^2\left\vert\nu_k\right\vert^2$.
 
 Then, with the definition of $\boldsymbol{\rm {d}}_{\varrho(k)}$ introduced, the sub-problem is rewritten as follows:
 \begin{subequations}\label{Problem:perfectB1}\small
 	\begin{alignat}{2}
 	\mathop{\max}\limits_{\boldsymbol{\rm {\beta}}}& \boldsymbol{\rm {\beta}}^H\hspace{-0.3mm}\boldsymbol{\rm {U}}_{R}\hspace{-0.3mm}\boldsymbol{\rm {\beta}}\hspace{-1mm}+\hspace{-1mm}\left(\hspace{-1.5mm}\sqrt{1\hspace{-1mm}-\hspace{-1mm}\boldsymbol{\rm {\beta}}^2}\hspace{-1mm}\right)^H\hspace{-2.5mm}\boldsymbol{\rm {U}}_{T}\sqrt{1\hspace{-1mm}-\hspace{-1mm}\boldsymbol{\rm {\beta}}^2}\hspace{-1mm}+\hspace{-1mm}2\mbox{Re}\hspace{-1mm}\left\{\hspace{-1mm}\boldsymbol{\rm {u}}_{R}^H\boldsymbol{\rm {\beta}}\hspace{-1mm}+\hspace{-1mm}\boldsymbol{\rm {u}}_{T}^H\hspace{-1mm}\sqrt{1\hspace{-1mm}-\hspace{-1mm}\boldsymbol{\rm {\beta}}^2}\hspace{-1mm}\right\} \notag \\
 	{\rm{s.t.}}:&1).\ \  0\le\beta_m\le 1, \forall m \in \mathcal{M};\\
 	&\hspace{-5mm}2).\  \boldsymbol{\rm {\beta}}^H\boldsymbol{\rm {U}}_{k}\boldsymbol{\rm {\beta}}+2\mbox{Re}\left\{\boldsymbol{\rm {u}}_{k}^H\boldsymbol{\rm {\beta}}\right\}\le c_{3,k}, \forall k \in \mathcal{K_R},\\
 	&\hspace{-5mm}3).\  \hspace{-1mm}\left(\hspace{-2mm}\sqrt{\hspace{-1mm}1\hspace{-1mm}-\hspace{-1mm}\boldsymbol{\rm {\beta}}^2}\right)^H\hspace{-3.5mm}\boldsymbol{\rm {U}}_{k}\sqrt{\hspace{-1mm}1\hspace{-1mm}-\hspace{-1mm}\boldsymbol{\rm {\beta}}^2}\hspace{-1mm}+\hspace{-1mm}2\mbox{Re}\left\{\boldsymbol{\rm {u}}_{k}^H\hspace{-1mm}\sqrt{\hspace{-1mm}1\hspace{-1mm}-\hspace{-1mm}\boldsymbol{\rm {\beta}}^2}\right\}\hspace{-1mm}\le\hspace{-1mm} c_{3,k},\forall k \hspace{-1mm}\in\hspace{-1mm} \mathcal{K_T},
 	\end{alignat}
 \end{subequations}
 It is clear that the non-convex term $\sqrt{1-\boldsymbol{\rm {\beta}}^2}$ in the objective function and (\ref{Problem:perfectB1}{c}) make subproblem tricky. To further simplify the problem, we intend to introduce the majorization-minimization (MM) algorithm \cite{MM1}. To begin with, we give the following theorem:
 \begin{theorem}\label{theoremMM}
 	Let $\boldsymbol{\rm {U}}\in\mathbb{C}^{M\times M}$ is a Hermitian matrix and $\boldsymbol{\rm {\breve U}}\in\mathbb{C}^{M\times M}$ is another Hermitian matrix such that $\boldsymbol{\rm {\breve U}}\succeq\boldsymbol{\rm {U}}$. For any given point $\boldsymbol{\rm {d}}^{(i)}\in\mathbb{C}^{M\times 1}$, the quadratic form $\boldsymbol{\rm {d}}^H\boldsymbol{\rm {U}}\boldsymbol{\rm {d}}$ can be upperbounded as
 	\begin{small}
 	\begin{align}
 		\boldsymbol{\rm {d}}^H\boldsymbol{\rm {U}}\boldsymbol{\rm {d}}\hspace{-1mm}\le\hspace{-1mm}\boldsymbol{\rm {d}}^H\boldsymbol{\rm {\breve U}}\boldsymbol{\rm {d}}\hspace{-1mm}+\hspace{-1mm}2\mbox{Re}\hspace{-1mm}\left\{\hspace{-1mm}\boldsymbol{\rm {d}}^H\hspace{-1mm}\left(\hspace{-1mm}\boldsymbol{\rm {U}}\hspace{-1mm}-\hspace{-1mm}\boldsymbol{\rm {\breve U}}\right)\hspace{-1mm}\boldsymbol{\rm {d}}^{(i)}\hspace{-1mm}\right\}\hspace{-1mm}+\hspace{-1mm}\boldsymbol{\rm {d}}^{(i)H}\hspace{-1mm}\left(\hspace{-1mm}\boldsymbol{\rm {\breve U}}-\boldsymbol{\rm { U}}\hspace{-1mm}\right)\hspace{-1mm}\boldsymbol{\rm {d}}^{(i)}.
 	\end{align}	
 	\end{small}
 \end{theorem}
 \textit{proof:} See  Lemma 1 in \cite{MM2}.$\hfill\blacksquare$
 
 Based on theorem~\ref{theoremMM}, we can obtain upper-bounds for the objective function and the constraints (\ref{Problem:perfectB1}{b}) and (\ref{Problem:perfectB1}{c}). Let $\{\boldsymbol{\rm {U}}_R, \boldsymbol{\rm {U}}_T, \boldsymbol{\rm {U}}_k\}$ refer to $\boldsymbol{\rm {U}}$ in the theorem~\ref{theoremMM}, as well as let $\{\lambda_{\max}(\boldsymbol{\rm {U}}_R)\boldsymbol{\rm {I}}_M, \lambda_{\max}(\boldsymbol{\rm {U}}_T)\boldsymbol{\rm {I}}_M, \lambda_{\max}(\boldsymbol{\rm {U}}_k)\boldsymbol{\rm {I}}_M\}$ belong to the corresponding $\boldsymbol{\rm {\breve U}}$, respectively, where $\lambda_{\max}(\cdot)$ denotes the maximum eigenvalue of the matrix. Therefore, the subproblem can be reconstructed as follows:
 \begin{subequations}\label{Problem:perfectB2}\small
 	\begin{alignat}{2}
 	\mathop{\max}\limits_{\boldsymbol{\rm {\beta}}}&
 	\lambda_{U_R}\left\Vert\boldsymbol{\rm {\beta}}\right\Vert^2_2+\lambda_{U_T}\left\Vert\sqrt{1\hspace{-1mm}-\hspace{-1mm}\boldsymbol{\rm{\beta}}^2}\right\Vert^2_2\hspace{-1mm}+\hspace{-1mm}2\mbox{Re}\hspace{-1mm}\left\{\hspace{-1mm}\boldsymbol{\rm {\widehat{u}}}^H_R\boldsymbol{\rm {\beta}}+\boldsymbol{\rm {\widehat{u}}}^H_T\sqrt{1\hspace{-1mm}-\hspace{-1mm}\boldsymbol{\rm {\beta}}^2}\right\}\notag\\
 	{\rm{s.t.}}:&1).\ \  0\le\beta_m\le 1, \forall m \in \mathcal{M};\\
 	&2).\  \lambda_{U_k}\left\Vert\boldsymbol{\rm {\beta}}\right\Vert^2_2+2\mbox{Re}\left\{\boldsymbol{\rm {\widehat{u}}}^H_{R_k}\boldsymbol{\rm {\beta}}\right\}\le \bar{c}_{3,k}, \forall k \in \mathcal{K_R},\\
 	&\hspace{-0.5mm}3). \lambda_{U_k}\hspace{-1mm}\left\Vert\sqrt{1\hspace{-1mm}-\hspace{-1mm}\boldsymbol{\rm {\beta}}^2}\right\Vert^2_2\hspace{-1mm}+\hspace{-1mm}2\mbox{Re}\left\{\boldsymbol{\rm {\widehat{u}}}^H_{T_k}\sqrt{1\hspace{-1mm}-\hspace{-1mm}\boldsymbol{\rm {\beta}}^2}\right\}\hspace{-1mm}\le\hspace{-1mm} \widehat{c}_{3,k}, \forall k \hspace{-1mm}\in\hspace{-1mm} \mathcal{K_T},
 	\end{alignat}
 \end{subequations}
 where the introduced parameters are defined as
 \begin{align}\label{Eqs:parameter1}
 &\lambda_{U_R}\hspace{-1mm}\triangleq\hspace{-1mm}\lambda_{\max}(\boldsymbol{\rm {U}}_R),\lambda_{U_T}\hspace{-1mm}\triangleq\lambda_{\max}(\boldsymbol{\rm {U}}_T),\lambda_{U_k}\hspace{-1mm}\triangleq\lambda_{\max}(\boldsymbol{\rm {U}}_k),\notag\\
 &\boldsymbol{\rm {\widehat{u}}}_R=\left(\boldsymbol{\rm {U}}_R-\lambda_{U_R}\boldsymbol{\rm {I}}_M\right)\boldsymbol{\rm {\beta}}^{(i)}+\boldsymbol{\rm {u}}_R,\notag\\
 &\boldsymbol{\rm {\widehat{u}}}_T=\left(\boldsymbol{\rm {U}}_T-\lambda_{U_T}\boldsymbol{\rm {I}}_M\right)\sqrt{1\hspace{-1mm}-\hspace{-1mm}{\boldsymbol{\rm {\beta}}^{(i)}}^2}+\boldsymbol{\rm {u}}_T,\notag\\
 &\boldsymbol{\rm {\widehat{u}}}_{R_k}=\left(\boldsymbol{\rm {U}}_k-\lambda_{U_k}\boldsymbol{\rm {I}}_M\right)\boldsymbol{\rm {\beta}}^{(i)}+\boldsymbol{\rm {u}}_k,\notag\\
 &\boldsymbol{\rm {\widehat{u}}}_{T_k}=\left(\boldsymbol{\rm {U}}_k-\lambda_{U_k}\boldsymbol{\rm {I}}_M\right)\sqrt{1\hspace{-1mm}-\hspace{-1mm}{\boldsymbol{\rm {\beta}}^{(i)}}^2}+\boldsymbol{\rm {u}}_k,\notag\\
 &\bar{c}_{3,k}=c_{3,k}-{\boldsymbol{\rm {\beta}}^{(i)}}^H\left(\lambda_{U_k}\boldsymbol{\rm {I}}_M-\boldsymbol{\rm {U}}_k\right)\boldsymbol{\rm {\beta}}^{(i)},\notag\\
 &\widehat{c}_{3,k}=c_{3,k}-\left({\sqrt{1\hspace{-1mm}-\hspace{-1mm}{\boldsymbol{\rm {\beta}}^{(i)}}^2}}\right)^H\hspace{-3mm}\left(\lambda_{U_k}\boldsymbol{\rm {I}}_M-\boldsymbol{\rm {U}}_k\right)\sqrt{1\hspace{-1mm}-\hspace{-1mm}{\boldsymbol{\rm {\beta}}^{(i)}}^2},
 \end{align}
 where $\boldsymbol{\rm {\beta}}^{(i)}$ represent the latest value of $\boldsymbol{\rm {\beta}}$ obtained in the last iteration.
 
 For brevity, the subproblem is further recast as follows:
  \begin{subequations}\label{Problem:perfectB3}\small
 	\begin{alignat}{2}
 	\textbf{\textit{P}1-(C):}&
 	\mathop{\max}\limits_{\boldsymbol{\rm {\beta}}}\sum_{m=1}^{M}\Big(
 	\lambda_{U_R}\beta_m^2+\lambda_{U_T}-\lambda_{U_T}\beta_m^2\notag\\
 	&\qquad\qquad\qquad+\left.2\mbox{Re}\left\{\widehat{u}^*_{R,m}\beta_m+\widehat{u}^*_{T,m}\sqrt{1\hspace{-1mm}-\beta_m^2}\right\}\right)\notag\\
 	{\rm{s.t.}}\hspace{-1mm}:&1).\ \  0\le\beta_m\le 1, \forall m \in \mathcal{M};\\
 	&\hspace{-2mm}2).\hspace{-1mm}\sum_{m=1}^{M}\hspace{-1mm}\left(\hspace{-1mm}\lambda_{U_k}\beta_m^2\hspace{-1mm}+\hspace{-1mm}2\mbox{Re}\left\{\widehat{u}^*_{R_k,m}\beta_m\hspace{-0.5mm}\right\}\hspace{-0.5mm}\right)\hspace{-1mm}\le\hspace{-1mm} \bar{c}_{3,k},\hspace{-1mm} \forall k \hspace{-1mm}\in \hspace{-1mm}\mathcal{K_R}\hspace{-0.5mm},\\
 	&\hspace{-2mm}3).\sum_{m=1}^{M}\hspace{-1mm}\left(\hspace{-1mm}\lambda_{U_k}-\hspace{-1mm}\lambda_{U_k}\beta_m^2+2\mbox{Re}\left\{\widehat{u}^*_{T_k,m}\sqrt{1\hspace{-1mm}-\hspace{-1mm}\beta_m^2}\right\}\hspace{-0.5mm}\right)\hspace{-1mm}\le \hspace{-1mm}\widehat{c}_{3,k},\notag \\ 
 	&\qquad\qquad\forall k \in \mathcal{K_T},
 	\end{alignat}
 \end{subequations}
  where the notations $\widehat{u}^*_{R,m}$, $\widehat{u}^*_{T,m}$, $\widehat{u}^*_{R_k,m}$, and $\widehat{u}^*_{T_k,m}$ represent the $m$th element of $\boldsymbol{\rm {\widehat{u}}}_{R}$, $\boldsymbol{\rm {\widehat{u}}}_{T}$, $\boldsymbol{\rm {\widehat{u}}}_{R_k}$, and $\boldsymbol{\rm {\widehat{u}}}_{T_k}$, respectively. Remarkably, we only have the tricky non-convex items $-\lambda_{U_T}\beta_m^2$ and $2\mbox{Re}\left\{\widehat{u}^*_{T,m}\sqrt{1\hspace{-1mm}-\hspace{-1mm}\beta_m^2}\right\}$ in the objective function and $-\lambda_{U_k}\beta_m^2$ and $2\mbox{Re}\left\{\widehat{u}^*_{T_k,m}\sqrt{1\hspace{-1mm}-\hspace{-1mm}\beta_m^2}\right\}$ in the constraint (\ref{Problem:perfectB3}{c}) left to deal with. To this end, we first investigate the convex surrogates of the concave terms $-\lambda_{U_T}\beta_m^2$ and $-\lambda_{U_k}\beta_m^2$, as follows:
  \begin{small}
  	\begin{align}\label{Eqs:convex1}
  		-\lambda_{U_T}\beta_m^2\le-\lambda_{U_T}{\beta_m^{(i)}}^2-2\lambda_{U_T}{\beta_m^{(i)}}^2\left(\beta_m^{(i)}-{\beta_m^{(i)}}^2\right),\\
  		-\lambda_{U_k}\beta_m^2\le-\lambda_{U_k}{\beta_m^{(i)}}^2-2\lambda_{U_k}{\beta_m^{(i)}}^2\left(\beta_m^{(i)}-{\beta_m^{(i)}}^2\right).\label{Eqs:convex2}
  	\end{align}
  \end{small}
  Convexifying $2\mbox{Re}\hspace{-1mm}\left\{\widehat{u}^*_{T,m}\sqrt{1\hspace{-1mm}-\hspace{-1mm}\beta_m^2}\right\}$ and $2\mbox{Re}\hspace{-1mm}\left\{\widehat{u}^*_{T_k,m}\sqrt{1\hspace{-1mm}-\hspace{-1mm}\beta_m^2}\right\}$ is more complicated, for which we have the following discussion: $2\mbox{Re}\hspace{-1mm}\left\{\widehat{u}^*_{T,m}\sqrt{1\hspace{-1mm}-\hspace{-1mm}\beta_m^2}\right\}$ and $2\mbox{Re}\hspace{-1mm}\left\{\widehat{u}^*_{T_k,m}\sqrt{1\hspace{-1mm}-\hspace{-1mm}\beta_m^2}\right\}$ are convex if $\widehat{u}^*_{T,m}\le 0$ and $\widehat{u}^*_{T_k,m}\le 0$, otherwise $2\mbox{Re}\hspace{-1mm}\left\{\widehat{u}^*_{T,m}\sqrt{1\hspace{-1mm}-\hspace{-1mm}\beta_m^2}\right\}$ and $2\mbox{Re}\hspace{-1mm}\left\{\widehat{u}^*_{T_k,m}\sqrt{1\hspace{-1mm}-\hspace{-1mm}\beta_m^2}\right\}$ has an upper-bound as follows:
  \begin{small}
  	\begin{align}\label{Eqs:convex3}
  		&\sqrt{1\hspace{-1mm}-\hspace{-1mm}\beta_m^2}\hspace{-1mm}\le\hspace{-1mm}-\hspace{-1mm}\left(\hspace{-1mm}\beta_m^{(i)}\sqrt{1\hspace{-1mm}-\hspace{-1mm}{\beta_m^{(i)}}^2}\right)\hspace{-1mm}\beta_m^2\hspace{-1mm}+\hspace{-1mm}\sqrt{1\hspace{-1mm}-\hspace{-1mm}{\beta_m^{(i)}}^2}\hspace{-1mm}+\hspace{-1mm}{\beta_m^{(i)}}^2\hspace{-1mm}\sqrt{1\hspace{-1mm}-\hspace{-1mm}{\beta_m^{(i)}}^2}^{-1}\hspace{-3mm},
  	\end{align}
  \end{small}

  Based on the above derivation, subproblem $\textbf{\textit{P}1-(C)}$ is constructed as a convex optimization problem, which can be solved with the convex optimization solver, such as CVX.
  
  \subsubsection{Update the phase shifts  $\boldsymbol{\rm {\theta}}_R$ and $\boldsymbol{\rm {\theta}}_T$}
  \label{subsubsec:theta}
  
  When other variables are given, we then investigate the update scheme for the phase shifts  $\boldsymbol{\rm {\theta}}_R$ and $\boldsymbol{\rm {\theta}}_T$. To begin with, we employ a similar approach to update $\boldsymbol{\rm {\beta}}$ to organize the objective function and constraints (\ref{constraint:12f}) of $\textbf{\textit{P}1-(A)}$ into the form with respect to $\boldsymbol{\rm {\theta}}_R$ and $\boldsymbol{\rm {\theta}}_T$ as follows ($\varrho\in\left\{R,T\right\}$):
  \begin{subequations}\label{Problem:perfectT}
  	\begin{alignat}{2}
  	\mathop{\min}\limits_{\boldsymbol{\rm {\theta}}_{\varrho}}& \ \sum_{\varrho\in\{R,T\}}\left(\boldsymbol{\rm {\theta}}_{\varrho}^H\boldsymbol{\rm {V}}_{\varrho}\boldsymbol{\rm {\theta}}_{\varrho}+2\mbox{Re}\left\{\boldsymbol{\rm {v}}_{\varrho}^H\boldsymbol{\rm {\theta}}_{\varrho}\right\}\right) \notag \\
  	{\rm{s.t.}}:&1).\ \left\vert\theta_{\varrho,m}\right\vert = 1, \forall m \in \mathcal{M};\\
  	&2).\  \boldsymbol{\rm {\theta}}_{\varrho}^H\boldsymbol{\rm {V}}_{k}\boldsymbol{\rm {\theta}}_{\varrho}+2\mbox{Re}\left\{\boldsymbol{\rm {v}}_{k}^H\boldsymbol{\rm {\theta}}_{\varrho}\right\}\le c_{4,k}, \forall k \in \mathcal{K},
  	\end{alignat}
  \end{subequations}
 where the above newly introduced definitions are given as follows:
 \begin{small}
 	\begin{align}
 	&\boldsymbol{\rm {V}}_{\varrho}\hspace{-0.5mm}\triangleq\hspace{-2mm}\sum_{k\in K_\varrho}\hspace{-2mm}\left\vert\nu_k\right\vert^2\hspace{-0.5mm}\sum_{j=1}^{K}\boldsymbol{\rm {\bar{\Upsilon}}}_{k,j}^*\boldsymbol{\rm {\bar{\Upsilon}}}_{k,j}^T, \boldsymbol{\rm {V}}_{k}\triangleq\left\vert\nu_k\right\vert^2\sum_{j=1}^{K}\boldsymbol{\rm {\bar{\Upsilon}}}_{k,j}^*\boldsymbol{\rm {\bar{\Upsilon}}}_{k,j}^T,\notag\\
 	&\boldsymbol{\rm {v}}_{\varrho}\triangleq\sum_{k\in K_\varrho}\left\vert\nu_k\right\vert^2\sum_{j=1}^{K}\kappa_{k,j}^*\boldsymbol{\rm {\bar{\Upsilon}}}_{k,j}^*\notag\\
 	&\qquad-\sum_{k\in K_\varrho}\left(\sqrt{1\hspace{-0.5mm}+\hspace{-0.5mm}\gamma_k}\nu_k^*\mbox{Diag}(\boldsymbol{\rm {h}}_{r,k}^*)\boldsymbol{\rm {D}}_{\varrho(k)}\boldsymbol{\rm {A}}\boldsymbol{\rm {G}}\boldsymbol{\rm {w}}_k\right)^*,\notag\\
 	&\boldsymbol{\rm {v}}_{k}\hspace{-1mm}\triangleq\hspace{-1mm}\left\vert\nu_k\right\vert^2\hspace{-1mm}\sum_{j=1}^{K}\kappa_{k,j}^*\hspace{-1mm}\boldsymbol{\rm {\bar{\Upsilon}}}_{k,j}^*\hspace{-1mm}-\hspace{-1mm}\hspace{-1mm}\left(\hspace{-1mm}\sqrt{1\hspace{-0.5mm}+\hspace{-0.5mm}\gamma_k}\nu_k^*\mbox{Diag}(\boldsymbol{\rm {h}}_{r,k}^*)\boldsymbol{\rm {D}}_{\varrho(k)}\boldsymbol{\rm {A}}\boldsymbol{\rm {G}}\boldsymbol{\rm {w}}_k\hspace{-1mm}\right)^*\hspace{-2mm},\notag\\
 	&c_{4,k} \hspace{-1mm}\triangleq-R_{\min}+\mbox{ln}(1+\gamma_k)-\gamma_k+c_{1,k}+c_{2,k}-\sigma_k^2\left\vert\nu_k\right\vert^2\nonumber\\
 	&\qquad\qquad\qquad-\sigma^2_{\scriptscriptstyle R\hspace{-0.2mm}I\hspace{-0.2mm}S}\left\vert\nu_k\right\vert^2\Vert\boldsymbol{\rm A}\boldsymbol{\rm D}_{\varrho(k)}\boldsymbol{\rm h}_{r,k}\Vert^2,\notag\\
 	&\boldsymbol{\rm {\bar{\Upsilon}}}_{k,j}\triangleq\mbox{Diag}(\boldsymbol{\rm {h}}_{r,k}^*)\boldsymbol{\rm {D}}_{\varrho(k)}\boldsymbol{\rm {A}}\boldsymbol{\rm {G}}\boldsymbol{\rm {w}}_j.
 	\end{align}
 \end{small}
 
 To handle the intricate constraints of the subproblem, the PDD framework \cite{PDD1} is considered. The PDD framework is a two-layer iterative process which alternates between updating the original and auxiliary variables in its inner layer and selectively updating the penalty coefficients and dual variables in its outer layer. First of all, regarding the constraints, we introduce auxiliary variables $\boldsymbol{\rm {\phi}}_{\varrho}$ and $\boldsymbol{\rm {\psi}}_{\varrho,k}, \forall k\in\mathcal{K}, \varrho\in\left\{R,T\right\}$ with the
 following equality constraints:
 \begin{small}
 	\begin{align}\label{PDD1}
 		&\boldsymbol{\rm {\phi}}_{\varrho} = \boldsymbol{\rm {\theta}}_{\varrho},\\
 		&\boldsymbol{\rm {\psi}}_{\varrho,k}= \boldsymbol{\rm {\theta}}_{\varrho}, \forall k\in\mathcal{K}\label{PDD2}
 	\end{align}
 \end{small}
 Hence, the subproblem is rewritten as follows:
 \begin{subequations}\label{Problem:perfectT1}\small
 	\begin{alignat}{2}
 	\mathop{\min}\limits_{\boldsymbol{\rm {\theta}}_{\varrho}, \boldsymbol{\rm {\phi}}_{\varrho}, \boldsymbol{\rm {\psi}}_{\varrho,k}}& \ \sum_{\varrho\in\{R,T\}}\left(\boldsymbol{\rm {\theta}}_{\varrho}^H\boldsymbol{\rm {V}}_{\varrho}\boldsymbol{\rm {\theta}}_{\varrho}+2\mbox{Re}\left\{\boldsymbol{\rm {v}}_{\varrho}^H\boldsymbol{\rm {\theta}}_{\varrho}\right\}\right) \notag \\
 	{\rm{s.t.}}:&1).\ \left\vert\phi_{\varrho,m}\right\vert = 1, \forall m \in \mathcal{M};\\
 	&2).\  \boldsymbol{\rm {\psi}}_{\varrho,k}^H\boldsymbol{\rm {V}}_{k}\boldsymbol{\rm {\psi}}_{\varrho,k}+2\mbox{Re}\left\{\boldsymbol{\rm {v}}_{k}^H\boldsymbol{\rm {\psi}}_{\varrho,k}\right\}\le c_{4,k}, \forall k \in \mathcal{K},\\
 	&3).\ (\ref{PDD1})~\text{and}~(\ref{PDD2}).\nonumber
 	\end{alignat}
 \end{subequations}
 By dualizing and penalizing constraints (\ref{PDD1}) and (\ref{PDD2}) into the objective function with dual variables $\{\boldsymbol{\rm {\mu}}_{\varrho,1}>0, \boldsymbol{\rm {\mu}}_{\varrho,2,k}>0, \forall k\in\mathcal{K}\}$  and a penalty factor $\rho$, we can obtain the augmented Lagrangian (AL) problem in the inner loop of the PDD framework, as follows:
 \begin{subequations}\label{Problem:perfectT2}\small
	\begin{alignat}{2}
	\textbf{\textit{P}1-(D):}&\notag\\
	\mathop{\min}\limits_{\boldsymbol{\rm {\theta}}_{\varrho}, \boldsymbol{\rm {\phi}}_{\varrho}, \boldsymbol{\rm {\psi}}_{\varrho,k}}&\hspace{-3mm}\sum_{\varrho\in\{R,T\}}\hspace{-2mm}\left(\hspace{-1mm}\boldsymbol{\rm {\theta}}_{\varrho}^H\boldsymbol{\rm {V}}_{\varrho}\boldsymbol{\rm {\theta}}_{\varrho}\hspace{-1mm}+\hspace{-1mm}2\mbox{Re}\left\{\boldsymbol{\rm {v}}_{\varrho}^H\boldsymbol{\rm {\theta}}_{\varrho}\right\}\hspace{-1mm}+\hspace{-1mm}\frac{1}{2\rho}\hspace{-1mm}\left\Vert\boldsymbol{\rm {\theta}}_{\varrho}\hspace{-1mm}-\hspace{-1mm}\boldsymbol{\rm {\phi}}_{\varrho}\hspace{-1mm}+\hspace{-1mm}\rho\boldsymbol{\rm {\mu}}_{\varrho,1}\right\Vert_2^2\right. \notag \\
	&\qquad\qquad\qquad+\left.\frac{1}{2\rho}\sum_{k=1}^{K}\left\Vert\boldsymbol{\rm {\theta}}_{\varrho}-\boldsymbol{\rm {\psi}}_{\varrho,k}+\rho\boldsymbol{\rm {\mu}}_{\varrho,2,k}\right\Vert_2^2\right)\notag\\
	{\rm{s.t.}}:&(\ref{Problem:perfectT1}{\text{a}})~\text{and}~(\ref{Problem:perfectT1}{\text{b}})\notag.
	\end{alignat}
 \end{subequations}

\begin{algorithm}[t]\small
	\caption{PDD Framework to Solve Problem \textbf{\textit{P}1-(D)}.} \label{alg:PDD}
	\begin{algorithmic}[1]
		\STATE Initialize $\boldsymbol{\rm {\theta}}_{\varrho}^{(0)}$,$\boldsymbol{\rm {\phi}}_{\varrho}^{(0)}$,$\boldsymbol{\rm {\psi}}_{\varrho,k}^{(0)}$,$\boldsymbol{\rm {\mu}}_{\varrho,1}^{(0)}$,$\boldsymbol{\rm {\mu}}_{\varrho,2,k}^{(0)}$,$\rho^{(0)}$, $\forall k\in\mathcal{K},\varrho\in\left\{R,T\right\}$, $\mathcal{S}_{\triangle}^{\min}=10^{-3}$, the accuracy $\mathcal{S}_{\triangle}^{out}=10^{-4}$, the number of iterations $\tau=0$, and the maximum number of iterations $\tau_{\max}=15$;
		\REPEAT
		\STATE Set $\boldsymbol{\rm {\theta}}_{\varrho}^{(\tau-1,0)}:=\boldsymbol{\rm {\theta}}_{\varrho}^{(\tau-1)}$,$\boldsymbol{\rm {\phi}}_{\varrho}^{(\tau-1,0)}:=\boldsymbol{\rm {\phi}}_{\varrho}^{(\tau-1)}$,$\boldsymbol{\rm {\psi}}_{\varrho,k}^{(\tau-1,0)}:=\boldsymbol{\rm {\psi}}_{\varrho,k}^{(\tau-1)}$, the number of iterations $q=0$, and the maximum number of iterations and $q_{\max}=15$;
		\REPEAT
		\STATE Update $\boldsymbol{\rm {\theta}}_{\varrho}^{(\tau-1,q+1)}$ and $\boldsymbol{\rm {\phi}}_{\varrho}^{(\tau-1,q+1)}$ by (\ref{theta_solve})	and (\ref{phi_solve});
		\STATE Update $\boldsymbol{\rm {\psi}}_{\varrho,k}^{(\tau-1,q+1)}$,$\forall k\in\mathcal{K}$ by (\ref{KKT1}) and $\mu_{\varrho,3,k}$,$\forall k\in\mathcal{K}$ according to the bisection search method, where details of the bisection search algorithm can be found in \cite{PAN-CUN};
		\STATE Update $q=q+1$;
		\UNTIL The objective function value converges or $q=q_{\max}$;
		\STATE Set $\boldsymbol{\rm {\theta}}_{\varrho}^{(\tau)}:=\boldsymbol{\rm {\theta}}_{\varrho}^{(\tau-1,\infty)}$,$\boldsymbol{\rm {\phi}}_{\varrho}^{(\tau)}:=\boldsymbol{\rm {\phi}}_{\varrho}^{(\tau-1,\infty)}$,$\boldsymbol{\rm {\psi}}_{\varrho,k}^{(\tau)}:=\boldsymbol{\rm {\psi}}_{\varrho,k}^{(\tau-1,\infty)}$;
		\IF{$\mathcal{S}_{\triangle}(i)\le\mathcal{S}_{\triangle}^{\min}$} 
		\STATE Update $\boldsymbol{\rm {\mu}}_{\varrho,1}^{(\tau+1)}$ and $\boldsymbol{\rm {\mu}}_{\varrho,2,k}^{(\tau+1)}$ by applying (\ref{canshu1}) and (\ref{canshu2}); 
		\STATE Update $\rho^{(\tau+1)}=\rho^{(\tau)}$;
		\ELSE
		\STATE Update $\boldsymbol{\rm {\mu}}_{\varrho,1}^{(\tau+1)}=\boldsymbol{\rm {\mu}}_{\varrho,1}^{(\tau)}$ and $\boldsymbol{\rm {\mu}}_{\varrho,2,k}^{(\tau+1)}=\boldsymbol{\rm {\mu}}_{\varrho,2,k}^{(\tau)}$;
		\STATE Update $\rho^{(\tau+1)}$ by applying (\ref{canshu3});
		\ENDIF
		\STATE Update $\tau= \tau+1$;
		\UNTIL $\mathcal{S}_{\triangle}\le\mathcal{S}_{\triangle}^{out}$ or $\tau=\tau_{\max}$.
	\end{algorithmic}
\end{algorithm}

 Based on the PDD framework, we reconstruct this AL problem into three subproblems for simplified solution. To begin with, with given $\boldsymbol{\rm {\phi}}_{\varrho}$ and $\boldsymbol{\rm {\psi}}_{\varrho,k}, \forall k\in\mathcal{K}$, the AL optimization problem with respect to $\boldsymbol{\rm {\theta}}_{\varrho}$ transforms into a simple unconstrained convex optimization problem. By letting the first-order derivative of the objective function of AL problem be 0, we can obtain the optimal phase shifts $\boldsymbol{\rm {\theta}}_{\varrho}$, as follows:
 \begin{small}
 	\begin{align}\label{theta_solve}
 		&\boldsymbol{\rm {\theta}}_{\varrho}^{\star}\hspace{-1mm}=\hspace{-1mm}\left(\frac{K\hspace{-1mm}+\hspace{-1mm}1}{\rho}\boldsymbol{\rm {I}}_{M}\hspace{-1mm}+\hspace{-1mm}\boldsymbol{\rm {V}}_{\varrho}\hspace{-1mm}\right)^{-1}\hspace{-1mm}\left(\hspace{-1mm}\frac{1}{\rho}\hspace{-1mm}\left(\hspace{-1mm}\boldsymbol{\rm {\phi}}_{\varrho}\hspace{-1mm}-\hspace{-1mm}\sum_{k=1}^{K}\hspace{-1mm}\boldsymbol{\rm {\psi}}_{\varrho,k}\hspace{-1mm}\right)\hspace{-1mm}-\hspace{-1mm}\left(\hspace{-1mm}\boldsymbol{\rm {\mu}}_{\varrho,1}\hspace{-1mm}+\hspace{-1mm}\sum_{k=1}^{K}\boldsymbol{\rm {\mu}}_{\varrho,2,k}\hspace{-1mm}\right)\hspace{-1mm}-\hspace{-1mm}2\boldsymbol{\rm {v}}_{\varrho}\hspace{-1mm}\right)\hspace{-1mm}.
 	\end{align}
 \end{small}

 In addition, with given $\boldsymbol{\rm {\theta}}_{\varrho}$ and $\boldsymbol{\rm {\psi}}_{\varrho,k}, \forall k\in\mathcal{K}$, by dropping the irrelevant
 terms with respect to $\boldsymbol{\rm {\phi}}_{\varrho}$ and utilizing $\left\Vert\boldsymbol{\rm {\phi}}_{\varrho}\right\Vert^2=M$, the AL optimization problem can be rewritten as follow:
 \begin{subequations}\label{Problem:perfectT3}\small
 	\begin{alignat}{2}
 	\mathop{\max}\limits_{\boldsymbol{\rm {\phi}}_{\varrho}}&
    \sum_{\varrho\in\{R,T\}}	\mbox{Re}\left\{\left(\boldsymbol{\rm {\theta}}_{\varrho}+\rho\boldsymbol{\rm {\mu}}_{\varrho,1}\right)^H\boldsymbol{\rm {\phi}}_{\varrho}\right\} ~{\rm{s.t.}}:&(\ref{Problem:perfectT1}{\text{a}}).\notag 
 	\end{alignat}
 \end{subequations}
 Thus, we can simply derive the closed-form solution $\boldsymbol{\rm {\phi}}_{\varrho}^{\star}$ as follows:
 \begin{small}
 	\begin{align}\label{phi_solve}
 		\boldsymbol{\rm {\phi}}_{\varrho}^{\star}=\mbox{exp}\left(j\arg\left(\boldsymbol{\rm {\theta}}_{\varrho}+\rho\boldsymbol{\rm {\mu}}_{\varrho,1}\right)\right).
 	\end{align}
 \end{small}
 Furthermore, when $\boldsymbol{\rm {\theta}}_{\varrho}$ and $\boldsymbol{\rm {\phi}}_{\varrho}$ are fixed, the subproblem to optimize $\boldsymbol{\rm {\psi}}_{\varrho,k}$ is described as follows:
 \begin{subequations}\label{Problem:perfectT4}
	\begin{alignat}{2}
	\mathop{\min}\limits_{\boldsymbol{\rm {\psi}}_{\varrho,k}}&
	\sum_{\varrho\in\{R,T\}}\left(\sum_{k=1}^{K}\left\Vert\boldsymbol{\rm {\theta}}_{\varrho}-\boldsymbol{\rm {\psi}}_{\varrho,k}+\rho\boldsymbol{\rm {\mu}}_{\varrho,2,k}\right\Vert_2^2\right) ~{\rm{s.t.}}:&(\ref{Problem:perfectT1}{\text{b}}).\notag 
	\end{alignat}
 \end{subequations}
 Then introducing the Lagrange multipliers $\mu_{\varrho,3,k}>0, \forall k\in\mathcal{K}$ associated with the constraint (\ref{Problem:perfectT1}{b}), the above subproblem naturally splits into $K$ independent unconstrained small problems, each of which is assigned as
 \begin{align}\label{AL:perfectT4}
 \mathcal{L}_k&\left(\boldsymbol{\rm {\psi}}_{\varrho,k}, \mu_{\varrho,3,k}\right)=\left\Vert\boldsymbol{\rm {\theta}}_{\varrho}-\boldsymbol{\rm {\psi}}_{\varrho,k}+\rho\boldsymbol{\rm {\mu}}_{\varrho,2,k}\right\Vert_2^2\nonumber\\
 &+\mu_{\varrho,3,k}\left(\boldsymbol{\rm {\psi}}_{\varrho,k}^H\boldsymbol{\rm {V}}_{k}\boldsymbol{\rm {\psi}}_{\varrho,k}+2\mbox{Re}\left\{\boldsymbol{\rm {v}}_{k}^H\boldsymbol{\rm {\psi}}_{\varrho,k}\right\}-c_{4,k}\right),
 \end{align}
 Based on the first-order optimality condition, the optimal solution of $\boldsymbol{\rm {\psi}}_{\varrho,k}$ can be obtained as follows:
 \begin{align}\label{KKT1}
 \boldsymbol{\rm {\psi}}_{\varrho,k}\hspace{-1mm}=\hspace{-1mm}\left(\boldsymbol{\rm {I}}_{M}+\mu_{\varrho,3,k}\boldsymbol{\rm {V}}_{k}\right)^{-1}\left(\boldsymbol{\rm {\theta}}_{\varrho}+\rho\boldsymbol{\rm {\mu}}_{\varrho,2,k}-\mu_{\varrho,3,k}\boldsymbol{\rm {v}}_{k}\right).
 \end{align}

 When choosing the $\mu_{\varrho,3,k}$, the complementary slackness conditions \cite{PAN-CUN} of the constraints (\ref{Problem:perfectT1}{b}) need to be satisfied as follows:
  \begin{align}\label{formula:KKT}
 \begin{split}
  \mu_{\varrho,3,k}\left(\boldsymbol{\rm {\psi}}_{\varrho,k}^H\boldsymbol{\rm {V}}_{k}\boldsymbol{\rm {\psi}}_{\varrho,k}+2\mbox{Re}\left\{\boldsymbol{\rm {v}}_{k}^H\boldsymbol{\rm {\psi}}_{\varrho,k}\right\}-c_{4,k}\right)=0.
  \end{split}
 \end{align}
 Thus, if the following condition holds:
 \begin{align}\label{formula:KKT1}
 \begin{split}
 \boldsymbol{\rm {\psi}}_{\varrho,k}^H\boldsymbol{\rm {V}}_{k}\boldsymbol{\rm {\psi}}_{\varrho,k}+2\mbox{Re}\left\{\boldsymbol{\rm {v}}_{k}^H\boldsymbol{\rm {\psi}}_{\varrho,k}\right\}\le c_{4,k},
 \end{split}
 \end{align}
 then the optimal solution for $\mu_{\varrho,3,k}$ is $\mu_{\varrho,3,k}^{\star}=0$. Otherwise, the following conditions are satisfied
 \begin{align}\label{formula:KKT2}
 \begin{split}
 \boldsymbol{\rm {\psi}}_{\varrho,k}^H\boldsymbol{\rm {V}}_{k}\boldsymbol{\rm {\psi}}_{\varrho,k}+2\mbox{Re}\left\{\boldsymbol{\rm {v}}_{k}^H\boldsymbol{\rm {\psi}}_{\varrho,k}\right\}= c_{4,k}.
\end{split}
 \end{align}
 Due to the complex structure of $\boldsymbol{\rm {\psi}}_{\varrho,k}^{\star}$, the closed-form solution for $\mu_{\varrho,3,k}$ is difficult to derive directly. To solve for $\mu_{\varrho,3,k}$, we additionally define the optimal solution of $\boldsymbol{\rm {\psi}}_{\varrho,k}$ by $\boldsymbol{\rm {\psi}}_{\varrho,k}\left(\mu_{\varrho,3,k}\right)$. Then, we have 
 \begin{align}
 \mathcal{P}\left(\mu_{\varrho,3,k}\right)\triangleq&\boldsymbol{\rm {\psi}}_{\varrho,k}^H\left(\mu_{\varrho,3,k}\right)\boldsymbol{\rm {V}}_{k}\boldsymbol{\rm {\psi}}_{\varrho,k}\left(\mu_{\varrho,3,k}\right)\notag\\
 &\qquad+2\mbox{Re}\left\{\boldsymbol{\rm {v}}_{k}^H\boldsymbol{\rm {\psi}}_{\varrho,k}\left(\mu_{\varrho,3,k}\right)\right\}= c_{4,k},
 \end{align}
 Finally, we can find $\mu_{\varrho,3,k}$ by solving the equation with respect to $\mathcal{P}\left(\mu_{\varrho,3,k}\right)$ by means of the bisection search method. For simplicity, the detailed solution procedure is omitted.
 
 As the inner layer of the PDD is an alternating iterative process, when its convergence is complete, the outer layer will choose to perform the following two operations. To begin with, we define that $\mathcal{S}_{\triangle}(i)=\max\left\{\left\Vert\boldsymbol{\rm {\theta}}_{\varrho}^{(i)}-\boldsymbol{\rm {\phi}}_{\varrho}^{(i)}\right\Vert_2,\left\Vert\boldsymbol{\rm {\theta}}_{\varrho}^{(i)}-\boldsymbol{\rm {\psi}}_{\varrho,1}^{(i)}\right\Vert_2,\cdots,\left\Vert\boldsymbol{\rm {\theta}}_{\varrho}^{(i)}-\boldsymbol{\rm {\psi}}_{\varrho,K}^{(i)}\right\Vert_2\right\}$. If $\mathcal{S}_{\triangle}(i)\le\mathcal{S}_{\triangle}^{\min}$, the dual variables $\boldsymbol{\rm {\mu}}_{\varrho,1}$ and $\boldsymbol{\rm {\mu}}_{\varrho,2,k}, \forall k\in\mathcal{K}$ will be updated according to the rule shown as follows:
 \begin{align}\label{canshu1}
 \boldsymbol{\rm {\mu}}_{\varrho,1}&=\boldsymbol{\rm {\mu}}_{\varrho,1}+\rho^{-1}\left(\boldsymbol{\rm {\theta}}_{\varrho}-\boldsymbol{\rm {\phi}}_{\varrho}\right),\\
 \boldsymbol{\rm {\mu}}_{\varrho,2,k}&=\boldsymbol{\rm {\mu}}_{\varrho,2,k}+\rho^{-1}\left(\boldsymbol{\rm {\theta}}_{\varrho}-\boldsymbol{\rm {\psi}}_{\varrho,k}\right),\forall k\in\mathcal{K},\label{canshu2}
 \end{align}
 where $\mathcal{S}_{\triangle}^{\min}$ denotes the tolerance of accuracy. Besides, if $\mathcal{S}_{\triangle}(i)>\mathcal{S}_{\triangle}^{\min}$, the outer layer will be chosen to increase the penalty parameter $\rho^{-1}$, which is updated according to the rule shown as follows:
 \begin{align}\label{canshu3}
 \rho^{-1}\leftarrow c_{\rho}^{-1}\rho^{-1},
 \end{align}
 where $c_{\rho}$ is a constant, which is typically chosen in the range of $[0.8, 0.9]$.

 The PDD-based algorithm for solving $\textbf{\textit{P}1-(D)}$ is summarized in Algorithm~\ref{alg:PDD}. Then, we investigate the complexity of the Algorithm~\ref{alg:PDD}. In each iteration, the main complexity is embodied in the calculation of $\boldsymbol{\rm {\theta}}_{\varrho}$ and $\boldsymbol{\rm {\psi}}_{\varrho,k}$,$\forall k\in\mathcal{K}$. The complexity of the calculation of $\boldsymbol{\rm {\theta}}_{\varrho}$ comes from solving for the inverse, thus as $\mathcal{O}\left(M^3\right)$. The total complexity of the calculation of $\boldsymbol{\rm {\psi}}_{\varrho,k}$,$\forall k\in\mathcal{K}$ is $\mathcal{O}\left(M^3\mbox{log}_2\frac{\Lambda_{\varrho,k}^{U}-\Lambda_{\varrho,k}^{L}}{2}\right)$, where $\Lambda_{\varrho,k}^{U}$ and $\Lambda_{\varrho,k}^{L}$ represent the the upper and lower bounds $\Lambda_{\varrho,k}$ in bisection search method. Therefore, the complexity of Algorithm~\ref{alg:PDD} is given by $\mathcal{O}_1 =\mathcal{O}\left(\tau_{\max}q_{\max}M^3\left(1+K\log_2\frac{\Lambda_{\varrho,k}^{U}-\Lambda_{\varrho,k}^{L}}{2}\right)\right)$.

 \subsubsection{Update the amplifier coefficient  $\boldsymbol{\rm {\alpha}}$}
 \label{subsubsec:alpha}
 
 We further investigate the optimization of the amplifier coefficient $\boldsymbol{\rm {\alpha}}$. When other variables are given, we again adopt a similar scheme for optimizing $\boldsymbol{\rm {\beta}}$ and first introduce the following variables:
 \begin{small}
 	\begin{align}
 	&\boldsymbol{\rm {\widetilde{\Upsilon}}}_{k,j}\triangleq\mbox{Diag}(\boldsymbol{\rm {h}}_{r,k}^*)\boldsymbol{\rm {\Theta}}_{\varrho(k)}\boldsymbol{\rm {D}}_{\varrho(k)}\boldsymbol{\rm {G}}\boldsymbol{\rm {w}}_j,\notag\\
 	&\boldsymbol{\rm {\bar{\Xi}}}\hspace{-0.5mm}\triangleq\hspace{-1mm}\sum_{k=1}^{K}\hspace{-1mm}\left\vert\nu_k\right\vert^2\hspace{-1mm}\left(\hspace{-1mm}\sum_{j=1}^{K}\boldsymbol{\rm {\widetilde{\Upsilon}}}_{k,j}\hspace{-1mm}\boldsymbol{\rm {\widetilde{\Upsilon}}}_{k,j}^H\hspace{-1mm}+\hspace{-1mm}\sigma^2_{\scriptscriptstyle R\hspace{-0.2mm}I\hspace{-0.2mm}S}\mbox{Diag}\left(\vert\boldsymbol{\rm {D}}_{\varrho(k)}\boldsymbol{\rm {h}}_{r,k}\vert^2\right)\hspace{-1mm}\right)\hspace{-1mm},\notag\\
 	&\boldsymbol{\rm {\epsilon}}\triangleq\sum_{k=1}^{K}\left\vert\nu_k\right\vert^2\sum_{j=1}^{K}\kappa_{k,j}\boldsymbol{\rm {\widetilde{\Upsilon}}}_{k,j}\notag\\
 	&\qquad-\sum_{k=1}^{K}\left(\sqrt{1\hspace{-0.5mm}+\hspace{-0.5mm}\gamma_k}\nu_k^*\mbox{Diag}(\boldsymbol{\rm {h}}_{r,k}^*)\boldsymbol{\rm {\Theta}}_{\varrho(k)}\boldsymbol{\rm {D}}_{\varrho(k)}\boldsymbol{\rm {G}}\boldsymbol{\rm {w}}_k\right),\notag\\
 	&\boldsymbol{\rm {\widetilde{\Xi}}}\hspace{-0.5mm}\triangleq\hspace{-0.5mm}\eta_{EE}\xi_{RIS}\left(\mbox{Diag}\left([\Vert\boldsymbol{\rm {g}}_1\Vert_2^2,\cdots,\Vert\boldsymbol{\rm {g}}_M\Vert_2^2]\right)\hspace{-1mm}+\hspace{-1mm}\sigma^2_{\scriptscriptstyle R\hspace{-0.2mm}I\hspace{-0.2mm}S}\boldsymbol{\rm {I}}_M\right),
 	\end{align}
 \end{small}
 where we introduce the notations $\boldsymbol{\rm {G}}\boldsymbol{\rm {W}}=\boldsymbol{\rm {G}}[\boldsymbol{\rm {w}}_1,\cdots,\boldsymbol{\rm {w}}_K]=[\boldsymbol{\rm {g}}_1,\cdots,\boldsymbol{\rm {g}}_M]^T$, where $\boldsymbol{\rm {g}}_m^T$ is the $m$th row of the matrix $\boldsymbol{\rm {G}}\boldsymbol{\rm {W}}$. Based on the above definition, we reorganize the objective function of $\textbf{\textit{P}1-(A)}$ into a function with respect to $\boldsymbol{\rm {\alpha}}$, which can be expressed as
 \begin{align}
 f_P^I&(\boldsymbol{\rm {\alpha}}; \boldsymbol{\rm {w}},\boldsymbol{\rm {\gamma}},\boldsymbol{\rm {\nu}},\boldsymbol{\rm {\beta}}, \boldsymbol{\rm {\theta}}_R,\boldsymbol{\rm {\theta}}_T)=\nonumber\\
 &~-\boldsymbol{\rm {\alpha}}^H\boldsymbol{\rm {\Xi}}\boldsymbol{\rm {\alpha}}-2\mbox{Re}\left\{\boldsymbol{\rm {\epsilon}}^H\boldsymbol{\rm {\alpha}}\right\}-\eta_{EE}\left\Vert\boldsymbol{\rm A}_c\right\Vert_0M_{RIS}P_{A}.
 \end{align}
 It can be seen that the main remaining difficulty to be addressed in the objective function arises from the term of $\left\Vert\boldsymbol{\rm A}_c\right\Vert_0$, which is non-convex and non-smooth.  In order to deal with $\left\Vert\boldsymbol{\rm A}_c\right\Vert_0$, a smooth approximation is applied to substitute the $\ell_0$-norm. Specifically, the $\ell_0$-norm can be replaced by the surrogate function \cite{l0norm} as follows:
 \begin{small}
 	\begin{align}\label{Eqs:A}
 		\left\Vert\boldsymbol{\rm A}_c\right\Vert_0\approx\frac{1}{\log(\frac{\delta+1}{\delta})}\sum_{l=1}^{L}\log\left(\frac{\delta+\left\Vert\boldsymbol{\rm \alpha}_l-\boldsymbol{\rm 1}_{M_{sub}}\right\Vert_2}{\delta}\right),
 	\end{align}
 \end{small}
 where $\delta$ represents a predefined positive constant, which is typically set with in the range of $[10^{-6}, 10^{-5}]$ to control the approximation precision. Then, we further transform Eq.~(\ref{Eqs:A}) by employing the MM method and obtain its convex upper-bound function $\hat{g}_{l}(\boldsymbol{\rm \alpha}|\boldsymbol{\rm \alpha}^{(i)})$ at the fixed point $\boldsymbol{\rm \alpha}^{(i)}$ as follows:
 \begin{small}
 	\begin{align}\label{alpha:norm}
 		&\log\hspace{-1mm}\left(\hspace{-1mm}\frac{\delta\hspace{-1mm}+\hspace{-1mm}\left\Vert\boldsymbol{\rm \alpha}_l\hspace{-1mm}-\hspace{-1mm}\boldsymbol{\rm 1}_{M_{sub}}\right\Vert_2}{\delta}\hspace{-1mm}\right)\le \hat{g}_{l}(\boldsymbol{\rm \alpha}|\boldsymbol{\rm \alpha}^{(i)})\nonumber\\
 		&=\log\hspace{-1mm}\left(\hspace{-1mm}\frac{\delta\hspace{-1mm}+\hspace{-1mm}\left\Vert\boldsymbol{\rm \alpha}^{(i)}_l\hspace{-1mm}-\hspace{-1mm}\boldsymbol{\rm 1}_{M_{sub}}\right\Vert_2}{\delta}\hspace{-1mm}\right)\hspace{-1mm}+\hspace{-1mm}\frac{\left\Vert\boldsymbol{\rm \alpha}_l\hspace{-1mm}-\hspace{-1mm}\boldsymbol{\rm 1}_{M_{sub}}\right\Vert_2\hspace{-1mm}-\hspace{-1mm}\left\Vert\boldsymbol{\rm \alpha}^{(i)}_l\hspace{-1mm}-\hspace{-1mm}\boldsymbol{\rm 1}_{M_{sub}}\right\Vert_2}{\left\Vert\boldsymbol{\rm \alpha}^{(i)}_l-\boldsymbol{\rm 1}_{M_{sub}}\right\Vert_2+\delta}\hspace{-1mm}.
 	\end{align}
 \end{small}

 Substituting $\hat{g}_{l}(\boldsymbol{\rm \alpha}|\boldsymbol{\rm \alpha}^{(i)})$ into the objective function, then we can obtain the following optimization subproblem:
 \begin{subequations}\label{Problem:perfectAlpha}\small
 	\begin{alignat}{2}
 	\mathop{\min}\limits_{\boldsymbol{\rm {\alpha}}}& \ \boldsymbol{\rm {\alpha}}^H\boldsymbol{\rm {\Xi}}\boldsymbol{\rm {\alpha}}+2\mbox{Re}\left\{\boldsymbol{\rm {\epsilon}}^H\boldsymbol{\rm {\alpha}}\right\}+\hat{g}(\boldsymbol{\rm \alpha}|\boldsymbol{\rm \alpha}^{(i)})\eta_{EE}M_{RIS}P_{A} \notag \\
 	{\rm{s.t.}}:&(\ref{Problem:perfect1}{\text{b}}), (\ref{Problem:perfect1}{\text{c}}), \text{and}\ (\ref{Problem:perfect1}{\text{f}}),\nonumber
 	\end{alignat}
 \end{subequations}
 where $\hat{g}(\boldsymbol{\rm \alpha}|\boldsymbol{\rm \alpha}^{(i)})={\log(\frac{\delta+1}{\delta})}^{-1}\sum_{l=1}^{L}\hat{g}_{l}(\boldsymbol{\rm \alpha}|\boldsymbol{\rm \alpha}^{(i)})$. Then, similar to the conversion of the objective function, the non-convex constraints (\ref{Problem:perfect1}{b}) and (\ref{Problem:perfect1}{f}) are modified into convex constraints, respectively, as follows:
 \begin{align}\label{const1}
&\boldsymbol{\rm {\alpha}}^H\boldsymbol{\rm {\widetilde{\Xi}}}\boldsymbol{\rm {\alpha}}+\hat{g}(\boldsymbol{\rm \alpha}|\boldsymbol{\rm \alpha}^{(i)})M_{RIS}P_{A}\le \bar{P}_{\max}^{RIS},\\
&\boldsymbol{\rm {\alpha}}^H\boldsymbol{\rm {\Xi}}_k\boldsymbol{\rm {\alpha}}+2\mbox{Re}\left\{\boldsymbol{\rm {\epsilon}}_k^H\boldsymbol{\rm {\alpha}}\right\}\le c_{5,k},\forall k \in\mathcal{K},\label{const2}
 \end{align}
 where we introduce the notations are defined as
 \begin{small}
 	\begin{align}
 	&\boldsymbol{\rm {\Xi}}_k\triangleq\hspace{-1mm}\left\vert\nu_k\right\vert^2\hspace{-1mm}\left(\hspace{-1mm}\sum_{j=1}^{K}\boldsymbol{\rm {\widetilde{\Upsilon}}}_{k,j}\hspace{-1mm}\boldsymbol{\rm {\widetilde{\Upsilon}}}_{k,j}^H\hspace{-1mm}+\hspace{-1mm}\sigma^2_{\scriptscriptstyle R\hspace{-0.2mm}I\hspace{-0.2mm}S}\mbox{Diag}\left(\vert\boldsymbol{\rm {D}}_{\varrho(k)}\boldsymbol{\rm {h}}_{r,k}\vert^2\right)\hspace{-1mm}\right)\hspace{-1mm},\notag\\
 	&\boldsymbol{\rm {\epsilon}}_k\triangleq\left\vert\nu_k\right\vert^2\sum_{j=1}^{K}\kappa_{k,j}\boldsymbol{\rm {\widetilde{\Upsilon}}}_{k,j}\notag\\
 	&\qquad\quad-\left(\sqrt{1\hspace{-0.5mm}+\hspace{-0.5mm}\gamma_k}\nu_k^*\mbox{Diag}(\boldsymbol{\rm {h}}_{r,k}^*)\boldsymbol{\rm {\Theta}}_{\varrho(k)}\boldsymbol{\rm {D}}_{\varrho(k)}\boldsymbol{\rm {G}}\boldsymbol{\rm {w}}_k\right)\hspace{-1mm},\notag\\
 	&c_{5,k}\hspace{-1mm}\triangleq\hspace{-1mm}-R_{\min}\hspace{-1mm}+\hspace{-1mm}\mbox{ln}(1\hspace{-0.5mm}+\hspace{-0.5mm}\gamma_k)\hspace{-1mm}-\gamma_k+c_{1,k}+c_{2,k}-\sigma_k^2\left\vert\nu_k\right\vert^2\hspace{-1mm},\notag\\
 	&\bar{P}_{\max}^{RIS}\triangleq P_{\max}^{RIS}-2MP_{PS}-MP_D.
 	\end{align}
 \end{small}
 
 After replacing the constraints, we can derive the following optimization problem:
 \begin{subequations}\label{Problem:perfectAlpha1}\small
 	\begin{alignat}{2}
 	\textbf{\textit{P}1-(E):}\mathop{\min}\limits_{\boldsymbol{\rm {\alpha}}}& \ \boldsymbol{\rm {\alpha}}^H\boldsymbol{\rm {\Xi}}\boldsymbol{\rm {\alpha}}+2\mbox{Re}\left\{\boldsymbol{\rm {\epsilon}}^H\boldsymbol{\rm {\alpha}}\right\}+\hat{g}(\boldsymbol{\rm \alpha}|\boldsymbol{\rm \alpha}^{(i)})\eta_{EE}M_{RIS}P_{A} \notag \\
 	{\rm{s.t.}}:&(\ref{const1}), (\ref{Problem:perfect1}{\text{c}}), \text{and}\ (\ref{const2}),\nonumber
 	\end{alignat}
 \end{subequations}
 which is convex and can be numerically solved by standard
 convex tools, e.g., CVX.
 
 \begin{algorithm}[t]\small
 	\caption{PDD-based AO algorithm for Problem \textbf{\textit{P}1}.} \label{alg:Framwork_all}
 	\begin{algorithmic}[1]
 		\STATE Initialize $\boldsymbol{\rm {w}}^{(0)}$,$\boldsymbol{\rm {\theta}}_R^{(0)}$,$\boldsymbol{\rm {\theta}}_T^{(0)}$,$\boldsymbol{\rm {\alpha}}^{(0)}$, $\boldsymbol{\rm {\beta}}^{(0)}$, the accuracy $\eta_{\varsigma}=10^{-3}$, the number of iterations $i=0$, and the maximum number of iterations $i_{\max}=15$;
 		\REPEAT
 		\STATE Update $\boldsymbol{\rm {\gamma}}^{(i+1)}$ and $\boldsymbol{\rm {\nu}}^{(i+1)}$ by (\ref{gamma_k}) and (\ref{nu_k}), respectively;
 		\STATE Solve the subproblem $\textbf{\textit{P}1-(B)}$ and obtain the optimal solution $\boldsymbol{\rm {w}}^{(i+1)}$;
 		\STATE Solve the subproblem $\textbf{\textit{P}1-(C)}$ and obtain the optimal solution $\boldsymbol{\rm {\beta}}^{(i+1)}$;
 		\STATE Update $\boldsymbol{\rm {\theta}}_R^{(i+1)}$ and $\boldsymbol{\rm {\theta}}_T^{(i+1)}$ by employing PDD framework in Algorithm~\ref{alg:PDD};
 		\STATE Solve the subproblem $\textbf{\textit{P}1-(E)}$ and obtain the optimal solution $\boldsymbol{\rm {\alpha}}^{(i+1)}$;
 		\STATE Update $i=i+1$;
 		\UNTIL $\left\vert \eta_{EE}^{(i+1)}-\eta_{EE}^{(i)}\right\vert/\eta_{EE}^{(i+1)}<\eta_{\varsigma}$ or $i = i_{\max}$.
 	\end{algorithmic}
 \end{algorithm}
 
 \subsection{Determine the RA operating pattern}
 \label{subsec:RA}
 
 Based on the obtained solution of $\boldsymbol{\rm {\alpha}}^{\star}$, we further determine the operating pattern of the RA in the sub-array. Intuitively, the smaller the magnitude $\Vert\boldsymbol{\rm {\alpha}}_l^{\star}\Vert_2$ of the sub-array, the less the contribution to the QoS requirements, which should be prioritized to be switched off. Motivated by this, $\Vert\boldsymbol{\rm {\alpha}}_l^{\star}\Vert_2,\forall l\in\mathcal{L}$ is ranked in ascending order resulting in a ``switching" level $\{\varpi_1,\cdots,\varpi_{L}\}$, where  the smaller $\varpi_l$ is, the higher the deactivation priority is. Therefore, the RA of the sub-array with the highest deactivation priority should be switched off. For the sake of illustration, let the set of sub-arrays with activated RAs be redefined as $\mathcal{L}_s(Q)=\{\varpi_{Q+1},\cdots,\varpi_{L}\}$. We therefore attempt to determine the largest $Q$ and thereby the size of the set $\mathcal{L}_s(Q)$.
 
 However, the EE is influenced in a non-monotonic manner by the number $Q$ of sub-arrays with activated RAs. In fact, both the denominator and numerator parts of the EE benefit from increasing with $Q$. Thus, we attempt to search for all $L$ possibilities, which yields a linear search complexity. To begin with, EE maximization is performed in the $\mathcal{L}_s = \mathcal{L}$ case. Afterwards, in each iteration, we deactivate the RAs of the subarrays with the highest deactivation priority in the current active set and optimize its EE. Finally, we choose the number of sub-arrays with activated RAs $Q$ associated with the maximum EE as our final RA activation scheme. With the determined RA activation scheme, the $\ell_0$-norm becomes a constant, thus we only need to obtain the final EE by applying Algorithm~\ref{alg:Framwork_all}.

 The complete PDD-based AO algorithm for solving problem $\textbf{\textit{P}1}$ is summarized in Algorithm~\ref{alg:Framwork_all}. Next we analyze the complexity of Algorithm~\ref{alg:Framwork_all}. After the derivation in Section~\ref{sec:Alternating_optimization_perfect_CSI}, the subproblems are reconstructed as convex optimization problems to be solved by the interior-point-method (IPM) based CVX toolbox, hence the complexity is primarily focused on the solution process in CVX toolbox. Finally, the complexity of the  overall Algorithm~\ref{alg:Framwork_all} is given as $\mathcal{O}\left(i_{\max}\left(N^{3.5}+2M^{3.5}+2\mathcal{O}_1\right)\right)$.

 \section{Proposed CSMM-Based Alternating Optimization In The Case Of Imperfect CSI}
 \label{sec:Alternating_optimization_imperfect_CSI}

 In this section, we aim to maximize the EE of the system in the case of imperfect CSI. To begin with, similar to the case of perfect CSI in section \ref{sec:Alternating_optimization_perfect_CSI}, we consider the equivalent transformation of problem $\textbf{\textit{P}2}$ into the following form by employing Dinkelbach algorithm and closed-form FP approach:
 \begin{subequations}\label{Problem:ImperfectA}\small
 	\begin{alignat}{2}
 	\textbf{\textit{P}2-(A):}\hspace{-1.5mm}
 	&\mathop{\min}\limits_{\boldsymbol{\rm {\theta}}_R,\boldsymbol{\rm {\theta}}_T,\boldsymbol{\rm {\alpha}},\boldsymbol{\rm {\beta}}}f_I^I(\boldsymbol{\rm {\theta}}_R,\boldsymbol{\rm {\theta}}_T,\boldsymbol{\rm {\alpha}},\boldsymbol{\rm {\beta}})=\mathbb{E}_{\vartheta}\left[g_I(\boldsymbol{\rm {\theta}}_R,\boldsymbol{\rm {\theta}}_T,\boldsymbol{\rm {\alpha}},\boldsymbol{\rm {\beta}};\vartheta)\right]   \notag \\
 	{\rm{s.t.}}:&1).\ (\ref{Problem:perfect1}{\text{c}}), (\ref{Problem:perfect1}{\text{d}}), (\ref{Problem:perfect1}{\text{e}}), \text{and}\ (\ref{Problem:Imperfect2}{\text{a}}), \notag\\
 	&2).\  \mathbb{E}_{\vartheta}\left[g_{P,k}(\boldsymbol{\rm {w}},\boldsymbol{\rm {\gamma}},\boldsymbol{\rm {\nu}},\boldsymbol{\rm {\theta}}_R,\boldsymbol{\rm {\theta}}_T,\boldsymbol{\rm {\alpha}},\boldsymbol{\rm {\beta}})\right]\hspace{-1mm}\ge\hspace{-1mm} R_{\min}, \forall k \in \mathcal{K},
 	\end{alignat}
 \end{subequations}
 where 
 \begin{small}
 	\begin{align}
 		g_I(\boldsymbol{\rm {\theta}}_R,\boldsymbol{\rm {\theta}}_T,\boldsymbol{\rm {\alpha}},\boldsymbol{\rm {\beta}};\vartheta)&=\mathop{\min}\limits_{\boldsymbol{\rm {w}},\boldsymbol{\rm {\gamma}},\boldsymbol{\rm {\nu}}} -f_P^I(\boldsymbol{\rm {w}},\boldsymbol{\rm {\gamma}},\boldsymbol{\rm {\nu}},\boldsymbol{\rm {\theta}}_R,\boldsymbol{\rm {\theta}}_T,\boldsymbol{\rm {\alpha}},\boldsymbol{\rm {\beta}};\vartheta)\nonumber\\
 		&{\rm{s.t.}}:\  (\ref{Problem:perfect1}{\text{a}}),(\ref{Problem:perfect1}{\text{b}}), \text{and}\ (\ref{constraint:12f}).\nonumber
 	\end{align}
 \end{small}
 Based on the solution scheme for $\boldsymbol{\rm {w}}$, $\boldsymbol{\rm {\gamma}}$, and $\boldsymbol{\rm {\nu}}$ in section \ref{sec:Alternating_optimization_perfect_CSI}, we can obtain a stationary solution of $g_I(\boldsymbol{\rm {\theta}}_R,\boldsymbol{\rm {\theta}}_T,\boldsymbol{\rm {\alpha}},\boldsymbol{\rm {\beta}};\vartheta)$, which is denoted as $\widetilde{g}_I(\boldsymbol{\rm {\theta}}_R,\boldsymbol{\rm {\theta}}_T,\boldsymbol{\rm {\alpha}},\boldsymbol{\rm {\beta}};\vartheta)$. Similarly, substituting the fixed solutions of $\boldsymbol{\rm {w}}$, $\boldsymbol{\rm {\gamma}}$, and $\boldsymbol{\rm {\nu}}$, the constrains (\ref{Problem:Imperfect2}{a}) and (\ref{Problem:ImperfectA}{a}) are reconstructed as follows:
 \begin{small}
 	 \begin{align}\label{constraint:IM1}
 		&\mathbb{E}_{\vartheta}\left[\widetilde{f}_{c1}(\boldsymbol{\rm {\alpha}};\vartheta)\right]\le P_{\max}^{RIS}, \\
 		&\mathbb{E}_{\vartheta}\left[\widetilde{g}_{P,k}(\boldsymbol{\rm {\theta}}_R,\boldsymbol{\rm {\theta}}_T,\boldsymbol{\rm {\alpha}},\boldsymbol{\rm {\beta}};\vartheta)\right]\ge R_{\min}, \forall k \in \mathcal{K},\label{constraint:IM2}
 	\end{align}
 \end{small}
 where $\widetilde{f}_{c1}(\boldsymbol{\rm {\alpha}};\vartheta)$ and $\widetilde{g}_{P,k}(\boldsymbol{\rm {\theta}}_R,\boldsymbol{\rm {\theta}}_T,\boldsymbol{\rm {\alpha}},\boldsymbol{\rm {\beta}};\vartheta)$ represent the rearranged ${f}_{c1}$ and $g_{P,k}(\boldsymbol{\rm {w}},\boldsymbol{\rm {\gamma}},\boldsymbol{\rm {\nu}},\boldsymbol{\rm {\theta}}_R,\boldsymbol{\rm {\theta}}_T,\boldsymbol{\rm {\alpha}},\boldsymbol{\rm {\beta}})$ functions by substituting $\boldsymbol{\rm {w}}$, $\boldsymbol{\rm {\gamma}}$, and $\boldsymbol{\rm {\nu}}$.
 
 Then, $\textbf{\textit{P}2-(A)}$ can be approximated as follows:
 \begin{subequations}\label{Problem:ImperfectB}\small
 	\begin{alignat}{2}
 	\textbf{\textit{P}2-(B):}
 	&\mathop{\min}\limits_{\boldsymbol{\rm {\theta}}_R,\boldsymbol{\rm {\theta}}_T,\boldsymbol{\rm {\alpha}},\boldsymbol{\rm {\beta}}}f_I^{I\hspace{-0.5mm}I}(\boldsymbol{\rm {\theta}}_R,\boldsymbol{\rm {\theta}}_T,\boldsymbol{\rm {\alpha}},\boldsymbol{\rm {\beta}})=\mathbb{E}_{\vartheta}\left[\widetilde{g}_I(\boldsymbol{\rm {\theta}}_R,\boldsymbol{\rm {\theta}}_T,\boldsymbol{\rm {\alpha}},\boldsymbol{\rm {\beta}};\vartheta)\right]   \notag \\
 	{\rm{s.t.}}:&\ (\ref{Problem:perfect1}{\text{c}}), (\ref{Problem:perfect1}{\text{d}}), (\ref{Problem:perfect1}{\text{e}}), (\ref{constraint:IM1}), \text{and}\ (\ref{constraint:IM2}).\notag
 	\end{alignat}
 \end{subequations}
 It is worth noting that when we acquire a stationary solution to the inner subproblem, the stationary solution to $\textbf{\textit{P}2-(B)}$ is also a stationary solution to $\textbf{\textit{P}2-(A)}$.
 
 \subsection{Constrained Stochastic Majorization-Minimization Algorithm for $\textbf{\textit{P}2-(B)}$}
 \label{Stochastic Majorization-Minimization}
 
 Since $\textbf{\textit{P}2-(B)}$ is non-convex, we focus on designing an efficient algorithm to find the stationary point of $\textbf{\textit{P}2-(B)}$. Compared to the optimization problem in the perfect CSI case, the main challenge in solving $\textbf{\textit{P}2-(B)}$ arises from the stochastic characteristics of the objective and constraint functions (i.e., it is difficult to accurately calculate the expectation). For the special case in which $\vartheta$ is a deterministic number, $\textbf{\textit{P}2-(B)}$ degenerates into a deterministic optimization problem, i.e., the optimization problem in the perfect CSI case. However, in the stochastic case, it is more challenging to design an algorithm to solve a problem concerning a stochastic non-convex objective function and constraints.
 
 The sample average approximation (SAA) method \cite{stochastic} is a traditional approach to solve the aforementioned problem. Specifically, at the $r$th iteration, we can update $\boldsymbol{\rm {\theta}}_R$, $\boldsymbol{\rm {\theta}}_T$, $\boldsymbol{\rm {\alpha}}$, $\boldsymbol{\rm {\beta}}$, and $\vartheta_r$ according to the following regularity:
 \begin{subequations}\label{Problem:ImperfectC}\small
 	\begin{alignat}{2}
 	&\hspace{-7mm}\{\boldsymbol{\rm {\theta}}_R,\boldsymbol{\rm {\theta}}_T,\boldsymbol{\rm {\alpha}},\boldsymbol{\rm {\beta}}\}\leftarrow\arg\min_{\boldsymbol{\rm {\theta}}_R,\boldsymbol{\rm {\theta}}_T,\boldsymbol{\rm {\alpha}},\boldsymbol{\rm {\beta}}}\frac{1}{r}\sum_{j=1}^r\widetilde{g}_I(\boldsymbol{\rm {\theta}}_R,\boldsymbol{\rm {\theta}}_T,\boldsymbol{\rm {\alpha}},\boldsymbol{\rm {\beta}};\vartheta_j)\notag\\
 	&{\rm{s.t.}}:\ 1).~(\ref{Problem:perfect1}{\text{c}}), (\ref{Problem:perfect1}{\text{d}}), \text{and}~ (\ref{Problem:perfect1}{\text{e}}), \notag\\
 	&~ 2).\frac{1}{r}\sum_{j=1}^r\widetilde{f}_{c1}(\boldsymbol{\rm {\alpha}};\vartheta_j)\le P_{\max}^{RIS};\\
 	&~3).\frac{1}{r}\sum_{j=1}^r\widetilde{g}_{P,k}(\boldsymbol{\rm {\theta}}_R,\boldsymbol{\rm {\theta}}_T,\boldsymbol{\rm {\alpha}},\boldsymbol{\rm {\beta}};\vartheta_j)\ge R_{\min},\forall k \in \mathcal{K}.
 	\end{alignat}
 \end{subequations}

 However, the SAA method has high computational complexity since each iteration computes the solution on the average of a large number of random samples. To overcome the above-mentioned difficulties, we employ the CSMM method \cite{CSMM}, where a suitable upper-bound approximation is devised for the objective and constraint functions at each iteration and for each channel realization. The solution for the $r$th iteration is determined by replacing the objective and constraint
 functions with their upper-bound convex surrogate functions. In addition, since the variables to be optimized are highly coupled, we employ the AO algorithm to further update them. Therefore, the variables are updated by solving a series of  CSMM subproblems at the $r$th iteration of the algorithm.
 
 \begin{algorithm}[t]\small
 	\caption{CSMM-Based AO Algorithm for Problem \textbf{\textit{P}2}.} \label{alg:Framwork_imperfect}
 	\begin{algorithmic}[1]
 		\STATE Channel realizations $\boldsymbol{\rm h}_{d,k}(\vartheta)$, $\boldsymbol{\rm h}_{r,k}(\vartheta)$, $\boldsymbol{\rm G}(\vartheta)$ for all $k$, and initialize $\boldsymbol{\rm {w}}^{(0)}$, $\imath=0$, the number of iterations $\imath=i=0$, and the maximum number of iterations $\imath_{\max}=i_{\max}=15$; 
 		\REPEAT
 		\REPEAT
 		\STATE Update $\boldsymbol{\rm {\gamma}}^{(\imath+1)}$ and $\boldsymbol{\rm {\nu}}^{(\imath+1)}$ by (\ref{gamma_k}) and (\ref{nu_k}), respectively;
 		\STATE Solve the subproblem $\textbf{\textit{P}1-(B)}$ and obtain the optimal solution $\boldsymbol{\rm {w}}^{(\imath+1)}$;
 		\STATE Update $\imath=\imath+1$;
 		\UNTIL The objective function value of $f_P^I$ converges or $\imath=\imath_{\max}$;
 		\STATE Initialize $\boldsymbol{\rm {\theta}}_R^{(0)},\boldsymbol{\rm {\theta}}_T^{(0)},\boldsymbol{\rm {\alpha}}^{(0)}$, $\boldsymbol{\rm {\beta}}^{(0)}$, the number of iterations $r=0$, and the maximum number of iterations $r_{\max}=15$;
 		\REPEAT 
 		\STATE Obtain new channel realizations $\boldsymbol{\rm h}_{d,k}(\vartheta_r)$, $\boldsymbol{\rm h}_{r,k}(\vartheta_r)$, $\boldsymbol{\rm G}(\vartheta_r)$ for all $k$;
 		\STATE Solve the subproblem $\textbf{\textit{P}2-(C)}$ and obtain the optimal solution $\boldsymbol{\rm {\beta}}^{(r+1)}$;
 		\STATE Update $\boldsymbol{\rm {\theta}}_R^{(r+1)}$ and $\boldsymbol{\rm {\theta}}_T^{(r+1)}$ by employing PDD framework;
 		\STATE Solve the subproblem $\textbf{\textit{P}2-(D)}$ and obtain the optimal solution $\boldsymbol{\rm {\alpha}}^{(r+1)}$;
 		\STATE Update $r=r+1$;
 		\UNTIL The objective function value of $\textbf{\textit{P}2-(B)}$ converges or $r=r_{\max}$;
 		\STATE Update $i=i+1$;
 		\UNTIL The objective function value of $\textbf{\textit{P}2}$ converges or $i=i_{\max}$.
 	\end{algorithmic}
 \end{algorithm}

 \subsection{Solve CSMM subproblems}
 \label{CSMM_subproblem}
 \subsubsection{Update the power splitting parameter $\boldsymbol{\rm {\beta}}$}
 \label{subsubsec:beta_imperfect}
 
 With other variables being given, the subproblem with respect to $\boldsymbol{\rm {\beta}}$ can be expressed as follows:
 \begin{subequations}\label{Problem:ImperfectBeta}
 	\begin{alignat}{2}
 	\min_{\boldsymbol{\rm {\beta}}}&\frac{1}{r}\sum_{j=1}^r\widehat{g}_I(\boldsymbol{\rm {\beta}};\boldsymbol{\rm {\beta}}^{(j-1)}, \vartheta_j)\notag\\
 	{\rm{s.t.}}:&\ 1).~0\le\beta_m\le 1, \forall m \in \mathcal{M};\\
 	&\ 2).~\frac{1}{r}\sum_{j=1}^r\widehat{g}_{P,k}(\boldsymbol{\rm {\beta}};\boldsymbol{\rm {\beta}}^{(j-1)}, \vartheta_j)\ge R_{\min}, \forall k \in \mathcal{K},
 	\end{alignat}
 \end{subequations}
 where $\widehat{g}_I(\boldsymbol{\rm {\beta}};\boldsymbol{\rm {\beta}}^{(j-1)}, \vartheta_j)$ and $\widehat{g}_{P,k}(\boldsymbol{\rm {\beta}};\boldsymbol{\rm {\beta}}^{(j-1)}, \vartheta_j)$ denote the surrogate functions associated to $\boldsymbol{\rm {\beta}}$, which are locally upper-bounds of the original functions $\widetilde{g}_I(\boldsymbol{\rm {\theta}}_R,\boldsymbol{\rm {\theta}}_T,\boldsymbol{\rm {\alpha}},\boldsymbol{\rm {\beta}};\vartheta_j)$ and $\widetilde{g}_{P,k}(\boldsymbol{\rm {w}},\boldsymbol{\rm {\gamma}},\boldsymbol{\rm {\nu}},\boldsymbol{\rm {\theta}}_R,\boldsymbol{\rm {\theta}}_T,\boldsymbol{\rm {\alpha}},\boldsymbol{\rm {\beta}})$ around the
 point $\boldsymbol{\rm {\beta}}^{(j-1)}$. To ensure convergence of the CSMM algorithm, the surrogate upper-bound requires a continuous differentiable strong convexity with uniformly bounded second order derivatives. Fortunately, the surrogate functions in subproblem $\textbf{\textit{P}1-(C)}$ with respect to $\boldsymbol{\rm {\beta}}$ in the case of perfect CSI fulfill the above requirements. Thus, following a similar derivation to $\textbf{\textit{P}1-(C)}$, we have
 \begin{subequations}\label{Problem:imperfectB1}\small
 	\begin{alignat}{2}
 	&\textbf{\textit{P}2-(C):}
 	\mathop{\max}\limits_{\boldsymbol{\rm {\beta}}}\frac{1}{r}\sum_{j=1}^r\sum_{m=1}^{M}\Big(
 	\lambda_{U_R}\beta_m^2+\lambda_{U_T}-\lambda_{U_T}\beta_m^2\notag\\
 	&\qquad\qquad\qquad+\left.2\mbox{Re}\left\{\widehat{u}^*_{R,m}\beta_m+\widehat{u}^*_{T,m}\sqrt{1\hspace{-1mm}-\beta_m^2}\right\}\right)\notag\\
 	&{\rm{s.t.}}:1).\ \  0\le\beta_m\le 1, \forall m \in \mathcal{M};\\
 	&\hspace{-1mm}2).\frac{1}{r}\sum_{j=1}^r\sum_{m=1}^{M}\hspace{-1mm}\left(\hspace{-1mm}\lambda_{U_k}\beta_m^2\hspace{-1mm}+\hspace{-1mm}2\mbox{Re}\left\{\widehat{u}^*_{R_k,m}\beta_m\hspace{-0.5mm}\right\}\hspace{-0.5mm}\right)\hspace{-1mm}\le\hspace{-1mm} \bar{c}_{3,k},\hspace{-1mm} \forall k \hspace{-1mm}\in \hspace{-1mm}\mathcal{K_R}\hspace{-0.5mm};\\
 	&\hspace{-1mm}3).\frac{1}{r}\hspace{-1mm}\sum_{j=1}^r\hspace{-1mm}\sum_{m=1}^{M}\hspace{-1mm}\left(\hspace{-1mm}\lambda_{U_k}\hspace{-1mm}-\hspace{-1mm}\lambda_{U_k}\beta_m^2\hspace{-1mm}+\hspace{-1mm}2\mbox{Re}\left\{\widehat{u}^*_{T_k,m}\sqrt{1\hspace{-1mm}-\hspace{-1mm}\beta_m^2}\right\}\hspace{-0.5mm}\right)\hspace{-1mm}\le \widehat{c}_{3,k},\hspace{-1mm}\forall k\hspace{-1mm} \in \hspace{-1mm}\mathcal{K_T},
 	\end{alignat}
 \end{subequations}
 where the parameters in $\textbf{\textit{P}2-(C)}$ are obtained by substituting the superscript $i$ for $j$ of the corresponding parameters in $\textbf{\textit{P}1-(C)}$, where specific definitions are omitted due to space constraints. Consistent with the perfect CSI case, $\textbf{\textit{P}2-(C)}$ can be solved by a common convex optimization solver after a sub-case discussion.

  \subsubsection{Update the phase shifts  $\boldsymbol{\rm {\theta}}_R$ and $\boldsymbol{\rm {\theta}}_T$}
 \label{subsubsec:theta_imperfect}
 
 For given other variables, the subproblem regarding $\boldsymbol{\rm {\theta}}_R$ and $\boldsymbol{\rm {\theta}}_T$ is rewritten as follows ($\varrho\in\left\{R,T\right\}$):
 \begin{subequations}\label{Problem:Imperfecttheta}\small
 	\begin{alignat}{2}
 	\min_{\boldsymbol{\rm {\theta}}_{\varrho}}&\frac{1}{r}\sum_{j=1}^r\widetilde{g}_I(\boldsymbol{\rm {\theta}}_{\varrho}; \vartheta_j)\notag\\
 	{\rm{s.t.}}:&\ 1).~\left\vert\theta_{\varrho,m}\right\vert = 1, \forall m \in \mathcal{M};\\
 	&\ 2).~\frac{1}{r}\sum_{j=1}^r\widetilde{g}_{P,k}(\boldsymbol{\rm {\theta}}_{\varrho};\vartheta_j)\ge R_{\min}, \forall k \in \mathcal{K},
 	\end{alignat}
 \end{subequations}
 where the reason for making the subproblem non-convex is the constraint (\ref{Problem:Imperfecttheta}{a}).

 In order to solve the above-mentioned tough non-convex constraint, we adopt the PDD framework to derive the solution, which is similar to the optimization scheme of $\boldsymbol{\rm {\theta}}_{\varrho}$ in the perfect CSI case. Due to the space limitation, the solution process based on the PDD framework is omitted here.
 
 \subsubsection{Update the amplifier coefficient  $\boldsymbol{\rm {\alpha}}$}
 \label{subsubsec:alpha_imperfect}
 When other variables are given, the subproblem with respect to $\boldsymbol{\rm {\alpha}}$ can be reformulated as follows:
 \begin{subequations}\label{Problem:ImperfectAlpha}\small
 	\begin{alignat}{2}
 	\min_{\boldsymbol{\rm {\beta}}}&\frac{1}{r}\sum_{j=1}^r\widehat{g}_I(\boldsymbol{\rm {\alpha}};\boldsymbol{\rm {\alpha}}^{(j-1)}, \vartheta_j)\notag\\
 	\hspace{-2mm}{\rm{s.t.}}:&\ 1).0\le\alpha_m\le\alpha_{\max}, \forall m \in \mathcal{M};\\
 	&\hspace{-2mm} 2).\frac{1}{r}\sum_{j=1}^r\widehat{f}_{c1}(\boldsymbol{\rm {\alpha}};\boldsymbol{\rm {\alpha}}^{(j-1)},\vartheta_j)\le P_{\max}^{RIS};\\
 	&\hspace{-2mm} 3).\frac{1}{r}\sum_{j=1}^r\widehat{g}_{P,k}(\boldsymbol{\rm {\alpha}};\boldsymbol{\rm {\alpha}}^{(j-1)}, \vartheta_j)\ge R_{\min}, \forall k \in \mathcal{K},
 	\end{alignat}
 \end{subequations}
 where $\widehat{f}_{c1}(\boldsymbol{\rm {\alpha}};\boldsymbol{\rm {\alpha}}^{(j-1)},\vartheta_j)$ and $\widehat{g}_{P,k}(\boldsymbol{\rm {\alpha}};\boldsymbol{\rm {\alpha}}^{(j-1)}, \vartheta_j)$ represent the surrogate functions based on the MM method concerning  $\boldsymbol{\rm {\alpha}}$, which are upper-bounds of $\widetilde{f}_{c1}(\boldsymbol{\rm {\alpha}};\vartheta_j)$ and $\widetilde{g}_{P,k}(\boldsymbol{\rm {w}},\boldsymbol{\rm {\gamma}},\boldsymbol{\rm {\nu}},\boldsymbol{\rm {\theta}}_R,\boldsymbol{\rm {\theta}}_T,\boldsymbol{\rm {\alpha}},\boldsymbol{\rm {\beta}})$ around the
 point $\boldsymbol{\rm {\alpha}}^{(j-1)}$. Similar to subsection~\ref{subsubsec:beta_imperfect}, we can directly and simply utilize the surrogate functions in the perfect CSI case for optimizing $\boldsymbol{\rm {\alpha}}$. Thus, the subproblem can be obtained as follows:
 \begin{subequations}\label{Problem:imperfectAlpha1}\small
 	\begin{alignat}{2}
 	\textbf{\textit{P}2-(D):}&\mathop{\min}\limits_{\boldsymbol{\rm {\alpha}}} \frac{1}{r}\sum_{j=1}^r\boldsymbol{\rm {\alpha}}^H\boldsymbol{\rm {\Xi}}\boldsymbol{\rm {\alpha}}+2\mbox{Re}\left\{\boldsymbol{\rm {\epsilon}}^H\boldsymbol{\rm {\alpha}}\right\}\notag \\
 	&\qquad\qquad\qquad\qquad+\hat{g}(\boldsymbol{\rm \alpha}|\boldsymbol{\rm \alpha}^{(j)})\eta_{EE}M_{RIS}P_{A} \notag \\
 	{\rm{s.t.}}:&\hspace{-1mm}\ 1).0\le\alpha_m\le\alpha_{\max}, \forall m \in \mathcal{M};\\
 	&\hspace{-5mm} 2).\frac{1}{r}\sum_{j=1}^r\boldsymbol{\rm {\alpha}}^H\boldsymbol{\rm {\widetilde{\Xi}}}\boldsymbol{\rm {\alpha}}+\hat{g}(\boldsymbol{\rm \alpha}|\boldsymbol{\rm \alpha}^{(j)})M_{RIS}P_{A}\le \bar{P}_{\max}^{RIS};\\
 	&\hspace{-5mm}3).\frac{1}{r}\sum_{j=1}^r\boldsymbol{\rm {\alpha}}^H\boldsymbol{\rm {\Xi}}_k\boldsymbol{\rm {\alpha}}+2\mbox{Re}\left\{\boldsymbol{\rm {\epsilon}}_k^H\boldsymbol{\rm {\alpha}}\right\}\le c_{5,k},\forall k \in\mathcal{K},
 	\end{alignat}
 \end{subequations}
 which can be efficiently solved via CVX. Again, we employ the same RA operating pattern scheme as in subsection~\ref{subsec:RA}, hence it is omitted here.

 The complete CSMM-based AO algorithm for solving problem $\textbf{\textit{P}2}$ is summarized in Algorithm~\ref{alg:Framwork_imperfect}. In accordance with the results of the complexity analysis of Algorithm~\ref{alg:Framwork_all}, the complexity of overall Algorithm~\ref{alg:Framwork_imperfect} is given as $\mathcal{O}\left(i_{\max}\left(\imath_{\max}N^{3.5}+r_{\max}\left(2M^{3.5}+2\mathcal{O}_1\right)\right)\right)$. The comparison shows that there is an increase in the complexity of the algorithm for the imperfect CSI case.

\section{Simulation Results and Analysis}
\label{sec:simulation}

\begin{figure}[t]
 	\vspace{-15pt}
 	\centering
 	\includegraphics[scale=0.065]{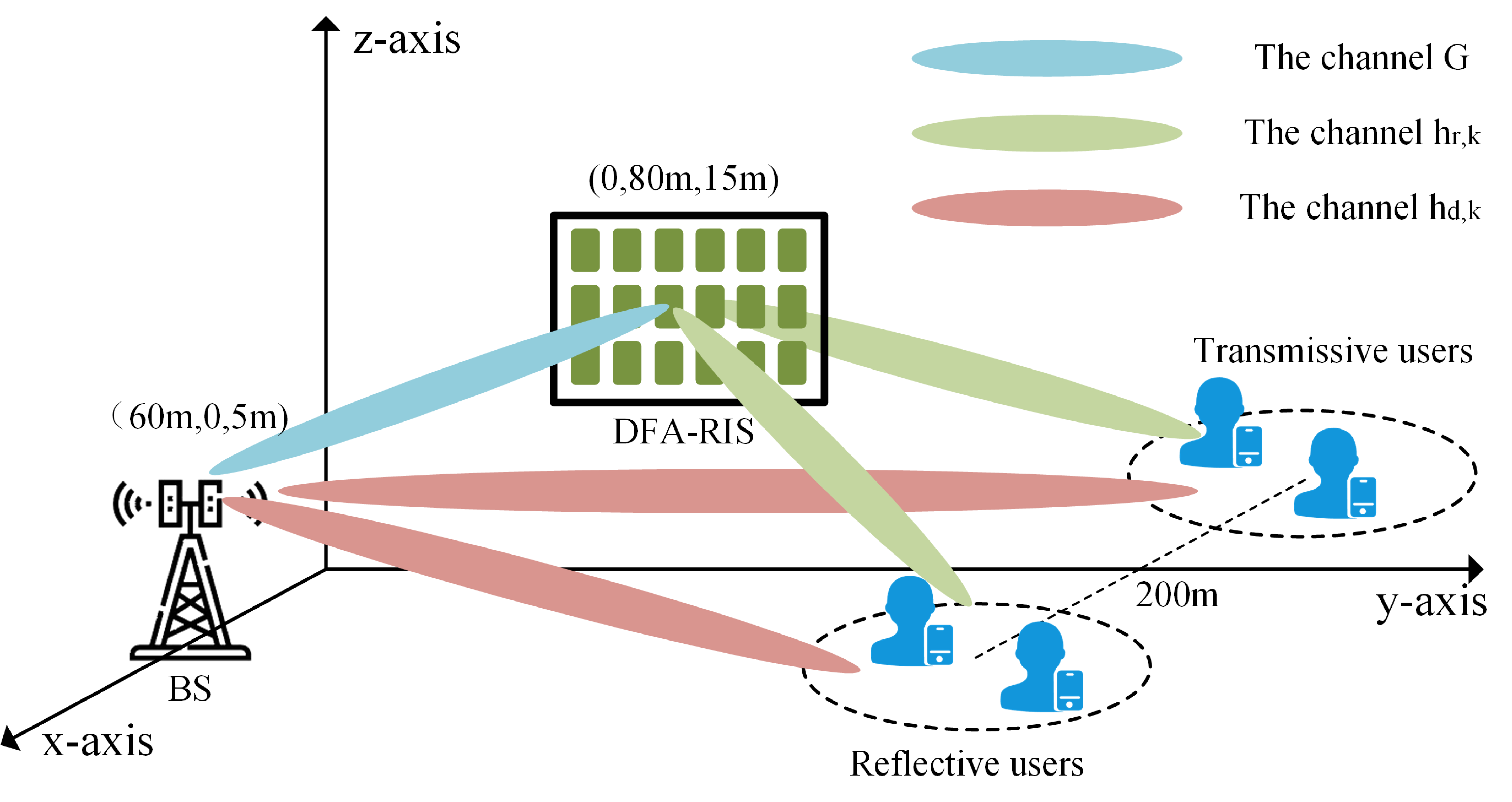}
 	\vspace{-10pt}
 	\caption{Plane diagram of simulated sub-array based DFA-RIS assisted wireless communication system.}
 	\vspace{-13pt}
 	\label{fig:system_model}
 \end{figure}

 In this section, we conduct numerical simulation experiments to evaluate the performance of our proposed dynamic energy-saving design of DFA-RIS in the multiuser MISO communication system. Unless otherwise stated, the setting of parameters run throughout our simulations. In the system, one BS equipped with $N=4$ antennas provides communication services to $K_R = 2$ reflective and $K_T = 2$ transmissive users with the assistance of the DFA-RIS based on sub-array. In the simulation, we divide $M = 60$ pairs of REs into $L=15$ sub-arrays. Each sub-array contains $M_{sub}=M/L=4$ pairs of REs. In order to facilitate the elaboration of the location of the devices in the system, we establish a coordinate system as in Fig.~\ref{fig:system_model}, where the coordinates of BS and DFA-RIS are (60m, 0, 5m) and (0, 80m, 15m), respectively. The reflective and transmissive users are located on circles of radius 15 centered at (30m, 200m, 0) and (-30m, 200m, 0), respectively. 
 
 We assume that the channels $\boldsymbol{\rm {G}}$ and $\boldsymbol{\rm {h}}_{r,k}$ follow the Rician distribution and the channel $\boldsymbol{\rm {h}}_{d,k}$ is a Rayleigh channel, where the Rician factor is $10$. The path loss of the wireless communication channel in dB is defined as $PL = \zeta_0 +10\chi_{pl} \mbox{log}_{10}(d/d_0)$, where $\zeta_0$ denotes the path loss at the reference distance $d_0=1$ m, and $\chi_{pl}$ represents the path loss exponent. In accordance with \cite{Joint-Active}, we adopt $\chi_{BU} = 3.45$, $\chi_{BR} = 2.2$, and $\chi_{RU} = 2.2$ to denote the path loss exponents of the channel between the BS and users, the channel between the BS and the DFA-RIS, and the channel between the DFA-RIS and users, respectively. In the case of imperfect CSI, we require estimates of the small-scale fading portion of the channels, i.e., the NLOS components of $\boldsymbol{\rm {G}}$ and $\boldsymbol{\rm {h}}_{r,k}$, and the entire $\boldsymbol{\rm {h}}_{d,k}$. Denote one of the above small-scale fadings as $x_h$, and the corresponding estimate as $\widehat{x}_h$. We further assume that the estimation error $x_h-\widehat{x}_h$ follows a zero-mean complex Gaussian distribution and all these small-scale fadings have the same normalized MSE, as $e_{MSE}=\mathbb{E}\left[\vert x_h-\widehat{x}_h\vert^2\right]/\mathbb{E}\left[\vert \widehat{x}_h\vert^2\right]$.
 
 In addition, we set the hardware static power of the BS, phase shifter, PDU, and RA as $P_{BS,c}=1.5$ W, $P_{PS}=1.25$ mW, $P_D=1.25$ mW, and $P_{A}=10$ mW,  respectively. The maximum available power consumption of DFA-RIS is $P_{\max}^{RIS}=30$ dBm. Furthermore, the energy conversion efficiencies of BS and DFA-RIS are both set to $1/1.1$, so $\xi_{BS}=\xi_{RIS}=1.1$. We set 
 the maximum amplified coefficient as $\alpha_{\max}=80$. For brevity, we assume that the noise power is the same for all users, i.e., $\sigma_k^2=\sigma^2=-100$ dBm, $\forall k \in\mathcal{K}$. The noise power introduced by the DFA-RIS is set to $\sigma^2_{\scriptscriptstyle R\hspace{-0.2mm}I\hspace{-0.2mm}S}= -100$ dBm. All simulation curves are obtained by averaging over 500 random tests.
 
 For comparative analysis, we compare our proposed dynamic energy-saving design of DFA-RIS with other RIS architectures, including fully-connected DFA-RIS, single-sided active-RIS (reflection and transmission) and passive STAR-RIS. Proposed sub-array based DFA-RIS, fully-connected DFA-RIS, and single-sided  active-RIS all have the same total transmitted and amplified power limit. For fairness, the transmit power budget in the system associated with the passive STAR-RIS architecture is given as the sum of the power budgets of the active-type RIS and BS.
 
 \begin{figure}[t]
 	\vspace{-15pt}
 	\centering
 	\includegraphics[scale=0.43]{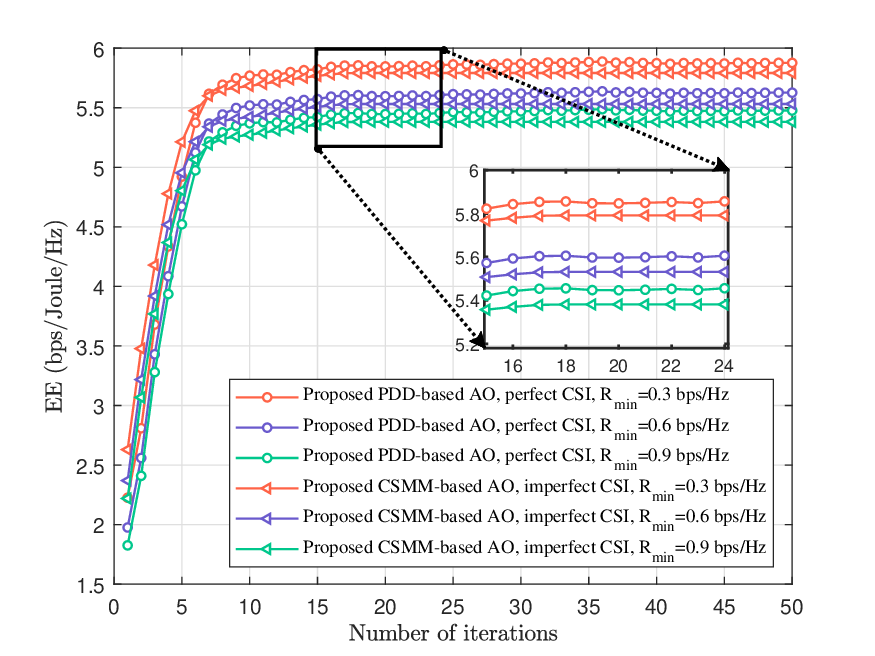}
 	\vspace{-10pt}
 	\caption{Energy efficiency versus the number of iterations.}
 	\vspace{-13pt}
 	\label{fig:ite_EE}
 \end{figure}
 
\begin{figure}[t]
	\centering
	\includegraphics[scale=0.43]{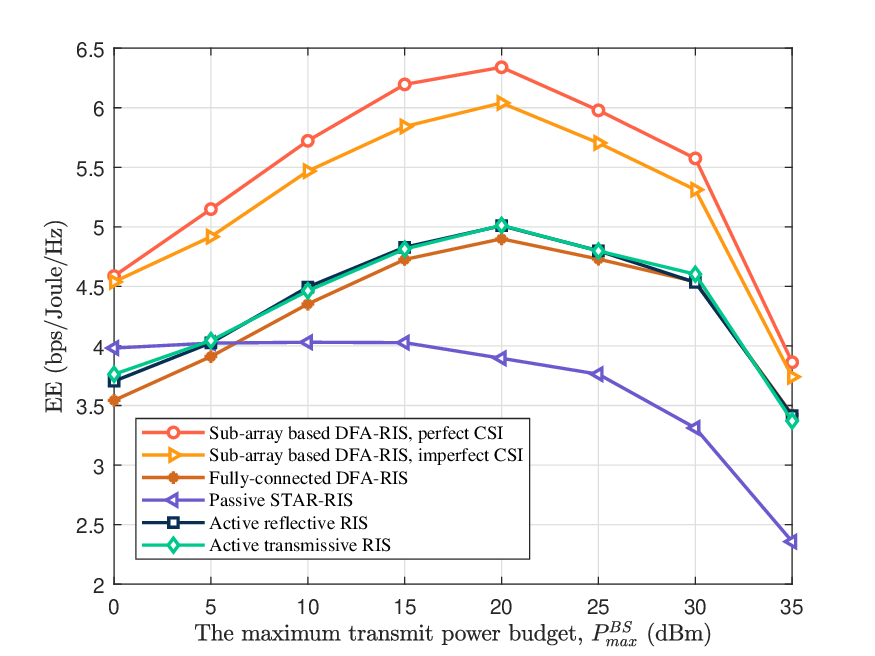}
	\vspace{-10pt}
	\caption{Energy efficiency versus the maximum transmit power budget $P_{\max}^{BS}$.}
	\vspace{-20pt}
	\label{fig:P_BS_EE}
\end{figure}
 
 \begin{figure}[t]
	\vspace{-15pt}
	\centering
	\includegraphics[scale=0.43]{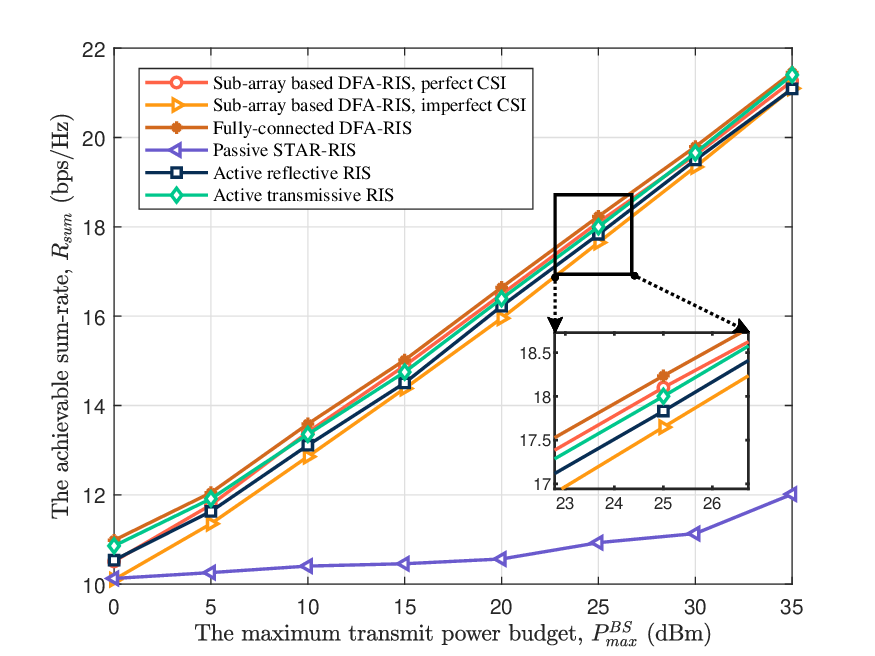}
	\vspace{-10pt}
	\caption{The achievable sum-rate  versus the maximum transmit power budget $P_{\max}^{BS}$.}
	\vspace{-20pt}
	\label{fig:P_BS_R}
\end{figure}
  The convergence behaviors  of Algorithm~\ref{alg:Framwork_all} and Algorithm~\ref{alg:Framwork_imperfect} are presented in Fig.~\ref{fig:ite_EE}. With the increase in the number of iterations, the EE of the system first significantly increases, and then keeps unchanged within $20$ iterations in the perfect/imperfect CSI case. In addition, EE decreases monotonically with increasing QoS requirements, indicating that a high QoS threshold further limits the EE of the system. This is due to the fact that more power is required to guarantee the increasing QoS requirements, resulting in an increase in power consumption and a decrease in EE.
 
  Figure~\ref{fig:P_BS_EE} shows the EE versus the maximum transmit power budget $P_{\max}^{BS}$ for different schemes. It can be seen that the EE of the system benefits more from the proposed dynamic energy-saving design of DFA-RIS than other competitors, where the EE in the imperfect CSI case is slightly inferior to the perfect CSI case. Fully-connected DFA-RIS and single-sided active-RIS have similar EE of the system and emerge as sub-optimal solutions due to the high power consumption. In addition, with increasing $P_{\max}^{BS}$, the EE of the schemes other than the passive STAR-RIS scheme first grows rapidly, peaking at around $P_{\max}^{BS}=20$ dBm. As $P_{\max}^{BS}$ continues to increase, the EE decreases significantly since the achievable rate of the system benefiting from $P_{\max}^{BS}$ struggles to compensate for the EE loss due to the increased power consumption. Furthermore, the EE of the passive STAR-RIS scheme remains consistently low and decreases as $P_{\max}^{BS}$ increases.

  Figure~\ref{fig:P_BS_R} depicts the achievable sum-rate performance $R_{sum}$ versus the maximum transmit power budget $P_{\max}^{BS}$. It can be seen that the sum-rate $R_{sum}$ increases
  with the growth of $P_{\max}^{BS}$. All active-type RISs provide significant performance gains over passive RIS since they are able to overcome double-fading attenuation. Combining Fig.~\ref{fig:P_BS_EE} and Fig.~\ref{fig:P_BS_R}, we can conclude that the proposed sub-array based DFA-RIS architecture can achieve similar sum-rate performance as the fully-connected DFA-RIS scheme with more substantial gains in the EE, which validates the advancement of sub-array based DFA-RIS architecture.

  \begin{figure}[t]
  	\centering
  	\includegraphics[scale=0.43]{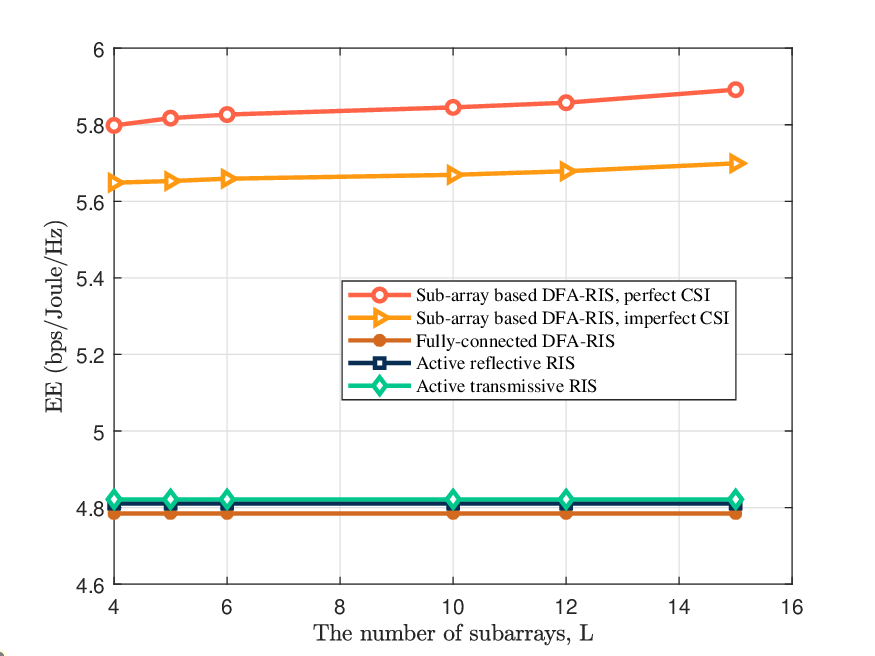}
  	\vspace{-10pt}
  	\caption{Energy efficiency versus the number of subarrays $L$.}
  	\vspace{-13pt}
  	\label{fig:L_EE}
  \end{figure}
  
 The effect of the number of sub-arrays on the EE of the system is shown in Fig.~\ref{fig:L_EE}. We still set the total number of RE pairs of the DFA-RIS at $60$. It can be seen that the higher the number of subarrays yields a higher EE. This is because a more accurate activation scheme for RAs can be performed, while introducing higher complexity. In addition, the EE of the systems associated with the remaining three active architectures does not change with increasing $L$.

 \begin{figure}[t]
 	\vspace{-15pt}
 	\centering
 	\includegraphics[scale=0.43]{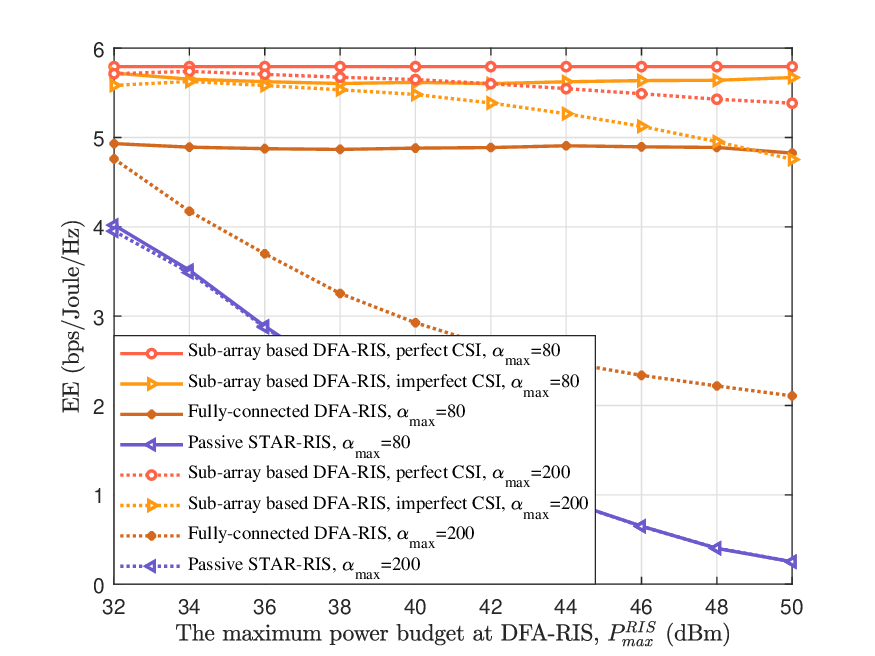}
 	\vspace{-10pt}
 	\caption{Energy efficiency versus the maximum power budget at DFA-RIS, $P_{\max}^{RIS}$.}
 	\vspace{-20pt}
 	\label{fig:P_RIS_max_EE}
 \end{figure}
 
 Figure~\ref{fig:P_RIS_max_EE} plots the EE of system versus the maximum power budget $P_{\max}^{RIS}$ at DFA-RIS using different schemes. For a given $\alpha_{\max} = 200$, the EE of the system for all schemes decreases with increasing $P_{\max}^{RIS}$. The reason is that the EE enhancement gained for larger $P_{\max}^{RIS}$ fails to offset the EE loss due to the increase in power consumption. Besides, it can be seen that our proposed dynamic energy-saving design gains less reduction from the increase in $P_{\max}^{RIS}$, compared to the other schemes. Interestingly, when $\alpha_{\max} = 80$, the EE obtained in the rest of the schemes except STAR-RIS remains constant as $P_{\max}^{RIS}$ increases. This interesting phenomenon is due to the fact that when $\alpha_{\max}$ is limited to smaller values, the constraint (\ref{Problem:perfect1}{b}) is always inactive in the optimization problem while the constraint (\ref{Problem:perfect1}{c}) is always active, indicating that the setting of $\alpha_{\max}$ is the dominant parameter affecting the EE.

 \begin{figure}[t]
 	\centering
 	\includegraphics[scale=0.45]{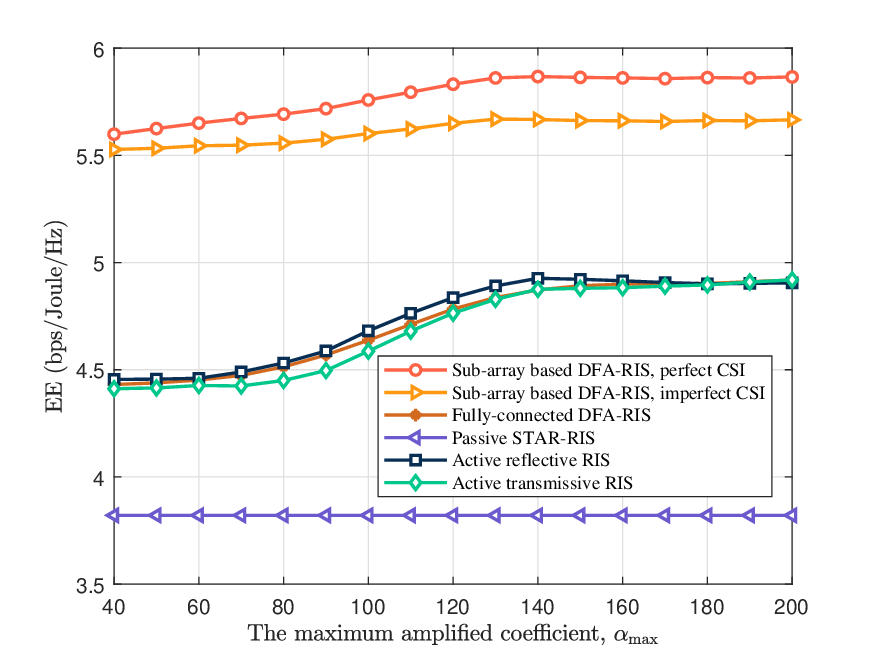}
 	\vspace{-10pt}
 	\caption{Energy efficiency versus the maximum amplified coefficient $\alpha_{\max}$.}
 	\vspace{-13pt}
 	\label{fig:alpha_max_EE}
 \end{figure}
 
 Figure~\ref{fig:alpha_max_EE} compares the EE versus the maximum amplified coefficient $\alpha_{\max}$. The growth of EE obtained in all active-type RISs gradually becomes moderate as $\alpha_{\max}$ increases. The reason is that for a given small $\alpha_{\max}$, the amplification of the active-type RIS is always bounded by $\alpha_{\max}$. However, when $\alpha_{\max}$ becomes sufficiently large, the constraint (\ref{Problem:perfect1}{c}) is deactivated and the EE is adversely impacted by other constraints.
\vspace{-5mm} 
\section{Conclusions}\label{sec:Conclusion}

 In this paper, we proposed a flexible and dynamic energy-saving design for DFA-RIS, namely the sub-array based DFA-RIS architecture, to reduce the power consumption and thus improve the EE of DFA-RIS based systems. In order to inspect the performance of this architecture in the system, we formulated joint the transmit beamforming,  DFA-RIS configuration, and RA operating pattern optimization problems to maximize the EE subject to minimum rate requirement, and power budgets at the BS and DFA-RIS. The PDD-based AO algorithm and the CSMM-based AO algorithm were efficiently developed to solve the nonconvex problem under perfect and imperfect CSI, respectively, aiming to maximize the EE of the system. Finally, the numerical results show that our proposed sub-array based DFA-RIS scheme yields higher EE compared to other RIS schemes.

%
 
\bibliographystyle{IEEEtran}
\bibliography{References}

\begin{thebibliography}{10}
\providecommand{\url}[1]{#1}
\csname url@samestyle\endcsname
\providecommand{\newblock}{\relax}
\providecommand{\bibinfo}[2]{#2}
\providecommand{\BIBentrySTDinterwordspacing}{\spaceskip=0pt\relax}
\providecommand{\BIBentryALTinterwordstretchfactor}{4}
\providecommand{\BIBentryALTinterwordspacing}{\spaceskip=\fontdimen2\font plus
\BIBentryALTinterwordstretchfactor\fontdimen3\font minus
  \fontdimen4\font\relax}
\providecommand{\BIBforeignlanguage}[2]{{%
\expandafter\ifx\csname l@#1\endcsname\relax
\typeout{** WARNING: IEEEtran.bst: No hyphenation pattern has been}%
\typeout{** loaded for the language `#1'. Using the pattern for}%
\typeout{** the default language instead.}%
\else
\language=\csname l@#1\endcsname
\fi
#2}}
\providecommand{\BIBdecl}{\relax}
\BIBdecl

\bibitem{zongshu1}
Z.~Wang, Y.~Du, K.~Wei, K.~Han, X.~Xu, G.~Wei, W.~Tong, P.~Zhu, J.~Ma, J.~Wang
  \emph{et~al.}, ``Vision, application scenarios, and key technology trends for
  {6G} mobile communications,'' \emph{Science China Information Sciences},
  vol.~65, no.~5, p. 151301, 2022.

\bibitem{OAM-NFC}
W.~Xu, Y.~Huang, W.~Wang, F.~Zhu, and X.~Ji, ``Toward ubiquitous and
  intelligent {6G} networks: from architecture to technology,'' \emph{Science
  China Information Sciences}, vol.~66, no.~3, p. 130300, 2023.

\bibitem{FullDuplex}
Z.~Yao, W.~Cheng, and H.~Zhang, ``Full-duplex assisted {LTE-U/Wifi} coexisting
  networks in unlicensed spectrum,'' \emph{IEEE Access}, vol.~6, pp.
  40\,085--40\,095, 2018.

\bibitem{Self-Interference}
J.~Wu, W.~Cheng, J.~Wang, J.~Wang, and W.~Zhang, ``{RIS}-based
  self-interference cancellation for full-duplex broadband transmission,''
  \emph{IEEE Transactions on Wireless Communications}, pp. 1--1, 2023.

\bibitem{Max_R}
M.-M. Zhao, Q.~Wu, M.-J. Zhao, and R.~Zhang, ``Intelligent reflecting surface
  enhanced wireless networks: {Two}-timescale beamforming optimization,''
  \emph{IEEE Transactions on Wireless Communications}, vol.~20, no.~1, pp.
  2--17, 2021.

\bibitem{Joint-Active}
Q.~Wu and R.~Zhang, ``Intelligent reflecting surface enhanced wireless network
  via joint active and passive beamforming,'' \emph{IEEE Transactions on
  Wireless Communications}, vol.~18, no.~11, pp. 5394--5409, 2019.

\bibitem{MISO-SEC}
Z.~Chu, W.~Hao, P.~Xiao, and J.~Shi, ``Intelligent reflecting surface aided
  multi-antenna secure transmission,'' \emph{IEEE Wireless Communications
  Letters}, vol.~9, no.~1, pp. 108--112, 2020.

\bibitem{Cao1}
Y.~Cao and W.~Cheng, ``Multiple reconfigurable intelligent surfaces assisted
  anti-jamming for aerial-ground communication,'' in \emph{ICC 2022 - IEEE
  International Conference on Communications}, 2022, pp. 698--703.

\bibitem{STAR_RIS1}
Y.~Liu, X.~Mu, J.~Xu, R.~Schober, Y.~Hao, H.~V. Poor, and L.~Hanzo, ``{STAR}:
  Simultaneous transmission and reflection for 360 coverage by intelligent
  surfaces,'' \emph{IEEE Wireless Communications}, vol.~28, no.~6, pp.
  102--109, 2021.

\bibitem{STAR_RIS2}
J.~Xu, Y.~Liu, X.~Mu, and O.~A. Dobre, ``{STAR-RISs}: Simultaneous transmitting
  and reflecting reconfigurable intelligent surfaces,'' \emph{IEEE
  Communications Letters}, vol.~25, no.~9, pp. 3134--3138, 2021.

\bibitem{Active_RIS1}
M.~H. Khoshafa, T.~M.~N. Ngatched, M.~H. Ahmed, and A.~R. Ndjiongue, ``Active
  reconfigurable intelligent surfaces-aided wireless communication system,''
  \emph{IEEE Communications Letters}, vol.~25, no.~11, pp. 3699--3703, 2021.

\bibitem{Active_RIS2}
Z.~Zhang, L.~Dai, X.~Chen, C.~Liu, F.~Yang, R.~Schober, and H.~V. Poor,
  ``Active ris vs. passive ris: Which will prevail in {6G}?'' \emph{IEEE
  Transactions on Communications}, vol.~71, no.~3, pp. 1707--1725, 2023.

\bibitem{DFA-RIS1}
Y.~Liu, Y.~Ma, M.~Li, Q.~Wu, and Q.~Shi, ``Spectral efficiency maximization for
  double-faced active reconfigurable intelligent surface,'' \emph{IEEE
  Transactions on Signal Processing}, vol.~70, pp. 5397--5412, 2022.

\bibitem{DFA-RIS2}
Y.~Guo, Y.~Liu, Q.~Wu, Q.~Shi, and Y.~Zhao, ``Enhanced secure communication via
  novel double-faced active {RIS},'' \emph{IEEE Transactions on
  Communications}, vol.~71, no.~6, pp. 3497--3512, 2023.

\bibitem{DFA-RIS3}
Y.~Ma, M.~Li, Y.~Liu, Q.~Wu, and Q.~Liu, ``Optimization for reflection and
  transmission dual-functional active {RIS}-assisted systems,'' \emph{IEEE
  Transactions on Communications}, vol.~71, no.~9, pp. 5534--5548, 2023.

\bibitem{Dinkelbach}
K.~Liu, Z.~Zhang, L.~Dai, S.~Xu, and F.~Yang, ``Active reconfigurable
  intelligent surface: Fully-connected or sub-connected?'' \emph{IEEE
  Communications Letters}, vol.~26, no.~1, pp. 167--171, 2022.

\bibitem{Quadratic_Transform}
K.~Shen and W.~Yu, ``Fractional programming for communication systems—{Part
  I}: Power control and beamforming,'' \emph{IEEE Transactions on Signal
  Processing}, vol.~66, no.~10, pp. 2616--2630, 2018.

\bibitem{CVX-intro}
M.~Grant and S.~Boyd, ``{CVX}: Matlab software for disciplined convex
  programming, version 2.1,'' \url{http://cvxr.com/cvx}, Mar. 2014.

\bibitem{MM1}
Y.~Sun, P.~Babu, and D.~P. Palomar, ``Majorization-minimization algorithms in
  signal processing, communications, and machine learning,'' \emph{IEEE
  Transactions on Signal Processing}, vol.~65, no.~3, pp. 794--816, 2017.

\bibitem{MM2}
J.~Song, P.~Babu, and D.~P. Palomar, ``Optimization methods for sequence design
  with low autocorrelation sidelobes,'' in \emph{2015 IEEE International
  Conference on Acoustics, Speech and Signal Processing (ICASSP)}, 2015, pp.
  3033--3037.

\bibitem{PDD1}
Q.~Shi and M.~Hong, ``Penalty dual decomposition method for nonsmooth nonconvex
  optimization—{Part I}: Algorithms and convergence analysis,'' \emph{IEEE
  Transactions on Signal Processing}, vol.~68, pp. 4108--4122, 2020.

\bibitem{PAN-CUN}
C.~Pan, H.~Ren, K.~Wang, M.~Elkashlan, A.~Nallanathan, J.~Wang, and L.~Hanzo,
  ``Intelligent reflecting surface aided {MIMO} broadcasting for simultaneous
  wireless information and power transfer,'' \emph{IEEE Journal on Selected
  Areas in Communications}, vol.~38, no.~8, pp. 1719--1734, 2020.

\bibitem{l0norm}
M.~Tao, E.~Chen, H.~Zhou, and W.~Yu, ``Content-centric sparse multicast
  beamforming for cache-enabled cloud ran,'' \emph{IEEE Transactions on
  Wireless Communications}, vol.~15, no.~9, pp. 6118--6131, 2016.

\bibitem{stochastic}
J.~Mairal, ``Stochastic majorization-minimization algorithms for large-scale
  optimization,'' \emph{Advances in Neural Information Processing Systems},
  vol.~26, 2013.

\bibitem{CSMM}
A.~Liu, V.~K.~N. Lau, and B.~Kananian, ``Stochastic successive convex
  approximation for non-convex constrained stochastic optimization,''
  \emph{IEEE Transactions on Signal Processing}, vol.~67, no.~16, pp.
  4189--4203, 2019.

\end{thebibliography}

\end{document}